\documentclass[USenglish]{article}

\usepackage{amssymb,amsmath}
\usepackage{graphicx}

\newtheorem{theorem}{Theorem}

\begin{document}

\author{Vanessa Robins \\ 
Department of Applied Mathematics\\ Research School of Physics and Engineering\\ The Australian National University\\ Canberra ACT 0200, Australia. \\  email: vanessa.robins@anu.edu.au }
\title{\textbf{Algebraic Topology}}
\maketitle

\section*{Abstract}
This manuscript will be published as Chapter 5 in Wiley's textbook \emph{Mathematical Tools for Physicists}, 2nd edition, edited by Michael Grinfeld from the University of Strathclyde. 
 
The chapter provides an introduction to the basic concepts of Algebraic Topology with an emphasis on  motivation from applications in the physical sciences.  It finishes with a brief review of computational work in algebraic topology, including persistent homology.



\clearpage

\tableofcontents

\clearpage

\section{Introduction}

Topology is the study of those aspects of shape and structure that do not depend on precise knowledge of an object's geometry. 
Accurate measurements are central to physics, so physicists like to joke that a topologist is someone who cannot tell the difference between a coffee cup and a doughnut.
However, the qualitative nature of topology and its ties to global analysis mean that many results are relevant to physical applications.  
One of the most notable areas of overlap comes from the study of dynamical systems.  
Some of the earliest work in algebraic topology was by Henri Poincar\'e in the 1890s, who pioneered a qualitative approach to the study of celestial mechanics by using topological results to prove the existence of periodic orbits \cite{Stillwell}.  
Topology continued to play an important role in dynamical systems with significant results pertinent to both areas from Smale in the 1960s \cite{Smale67}.  
More recently, in the 1990s computer analysis of chaotic dynamics was one of the drivers for innovation in computational topology \cite{Mindlin91,Muldoon93,Robins98,kaczynskibook}.

As with any established subject, there are several branches to topology:  \textbf{General topology} defines the notion of `closeness' (the neighborhood of a point), limits, continuity of functions and so on in the absence of a metric.  
These concepts are absolutely fundamental to modern functional analysis; a standard introductory  reference is \cite{Armstrong}.  
\textbf{Algebraic topology} derives algebraic objects (typically groups) from topological spaces to help determine when two spaces are alike.  
It also allows us to compute quantities such as the number of pieces the space has, and the number and type of `holes'.  
\textbf{Differential topology} builds on the above and on the differential geometry of manifolds to study the restrictions on functions that arise as the result of the structure of their domain.  
This chapter is primarily concerned with algebraic topology; it covers the elementary tools and concepts from this field.   
It draws on definitions and material from Chapter~\ref{GroupTheory} on group theory and Chapter~\ref{DifferentialGeometry} on differential geometry.

A central question in topology is to decide when two objects are the same in some sense.   
In general topology, two spaces, $A$ and $B$, are considered to be the same if there is a \textbf{homeomorphism}, $f$, between them: $f : A \to B$ is a continuous function with a continuous inverse.  
This captures an \emph{intrinsic} type of equivalence that allows arbitrary stretching and squeezing of a shape and permits changes in the way an object sits in a larger space (its embedding), but excludes any cutting or gluing.  
So for example, a circle ($x^2+y^2 = 1$) is homeomorphic to the perimeter of a square and to the trefoil knot, but not to a line segment, and a sphere with a single point removed is homeomorphic to the plane. 
One of the ultimate goals in topology is to find a set of quantities (called \emph{invariants}) that characterize spaces up to homeomorphism.
For arbitrary topological spaces this is known to be impossible~\cite{Stillwell} but  
for closed, compact 2-manifolds this problem is solved by finding the \emph{Euler characteristic} (see p.~\pageref{eulerchar}) and orientability of the surface \cite{SeifertThrelfall,ConwayZIP}.
     
What is the essential difference between a line segment and a circle? 
Intuitively it is the ability to trace your finger round and round the circle as many times as you like without stopping or turning back. 
Algebraic topology is the mathematical machinery that lets us quantify and detect this.  
The idea behind algebraic topology is to map topological spaces into groups (or other algebraic structures) in such a way that continuous functions between topological spaces map to homomorphisms between their associated groups.\footnote{%
A \textbf{homomorphism} between two groups, $\phi: G \to H$  is a function that respects the group operation. That is, $\phi(a\cdot b) = \phi(a)*\phi(b)$, for $a, b \in G$ where $\cdot$ is the group operation in $G$ and $*$ is the group operation in $H$.   
}


In Sections \ref{Homotopy}, \ref{Homology} and \ref{Cohomology}, this chapter covers the three basic constructions of algebraic topology: homotopy, homology and cohomology theories.  Each has a different method for defining a group from the structures in a topological space, and although there are close links between the three, they capture different qualities of a space.  
Many of the more advanced topics in algebraic topology involve studying functions on a space, so we introduce the fundamental link between critical points of a function and the topology of its domain in Section \ref{MorseTheory} on Morse Theory.  
The computability of invariants, both analytically and numerically, is vital to physical applications so the recent literature on computational topology is reviewed in Section \ref{ComputationalTopology}. 
Finally, we give a brief guide to further reading on applications of topology to physics.

\section{Homotopy Theory}
\label{Homotopy}

A homotopy equivalence is a weaker form of equivalence between topological spaces than a homeomorphism that allows us to collapse a space onto a lower-dimensional subset of itself (as we will explain in Section~\ref{HomotopyOfSpaces}), but it captures many essential aspects of shape and structure.  
When applied to paths in a space, homotopy equivalence allows us to define an algebraic operation on loops, and provides our first bridge between topology and groups.

\subsection{Homotopy of paths}

We begin by defining a \textbf{homotopy} between two continuous functions 
$f, g : X \to Y$.  
These maps will be homotopic if their images $f(X), \, g(X)$, can be continuously morphed from one to the other within $Y$, i.e.~there is a parametrized set of images that starts with one and ends with the other.  
Formally this deformation is achieved by defining a continuous function $F: X \times [0,1] \to Y$ 
with $F(x,0) = f(x)$ and $F(x,1) = g(x)$. 

\begin{figure}
\centering
\includegraphics[width=0.3\textwidth]{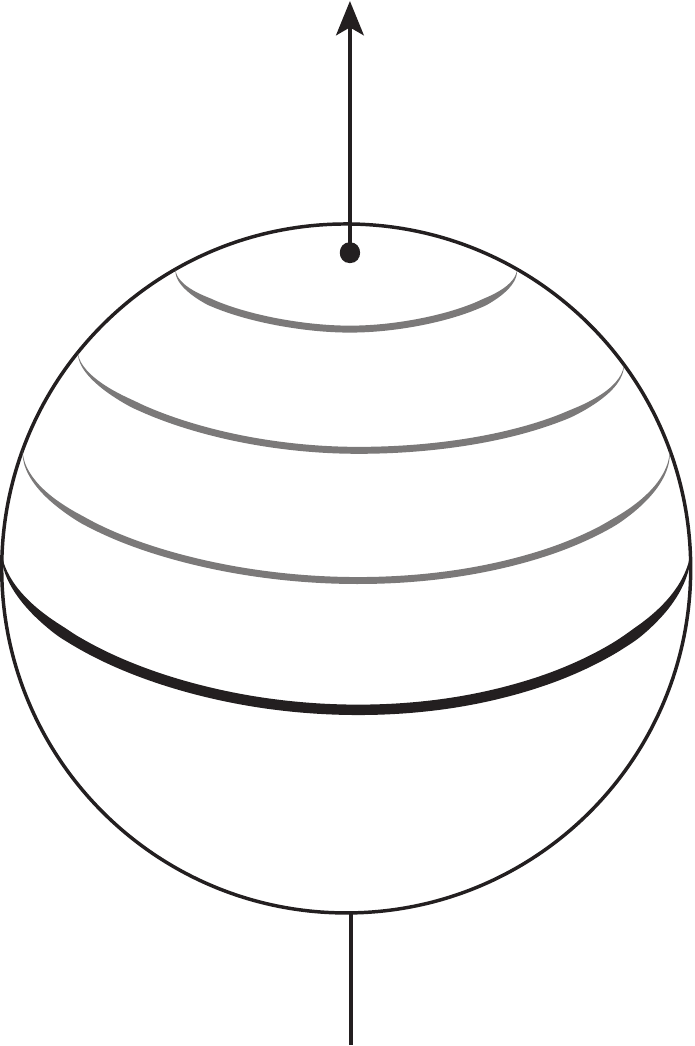}
\caption{The function $f: S^1 \to S^2$ that maps the circle onto the equator of the sphere is homotopic to the function $g: S^1 \to S^2$ that maps the circle to the north pole.  Three sections of the homotopy $F: S^1\times [0,1] \to S^2$ are shown in grey.}
\label{fig:sphere_homotopy}
\end{figure}

For example, consider two maps from the unit circle into the unit sphere, $f,g: S^1 \to S^2$.  
We use the angle $\theta \in [0,2\pi) $ to parametrize $S^1$ and the embedding $x^2 + y^2 + z^2 = 1$ in $\mathbb{R}^3$ to define points in $S^2$.   
Define $f(\theta) = (\cos \theta, \sin \theta, 0) $ to be a map from the circle to the equator and $g(\theta) = (0,0,1)$, a constant map from the circle to the north pole. 
A homotopy between $f$ and $g$ is given by $F(\theta, t) = (\cos(\pi t/2) \cos \theta, \cos(\pi t/2) \sin \theta, \sin(\pi t/2) ) $ and illustrated in Fig.~\ref{fig:sphere_homotopy}.
Any function that is homotopic to a constant function, as in this example, is called \textbf{null-homotopic}. 

When the domain is the unit interval, $X = [0,1]$, and $Y$ is an arbitrary topological space the functions $f$ and $g$ are referred to as \textbf{paths} in $Y$.  
A space in which every pair of points may be joined by a path is called \textbf{path-connected}.
It is often useful to consider homotopies between paths that fix their endpoints, $y_0$ and $y_1$, say, so we have the additional conditions on $F$ that for all $t \in [0,1]$, $F(0,t) = f(0) = g(0) = y_0$, and $F(1,t) = f(1) = g(1) = y_1$.  
If a path starts and ends at the same point, $y_0 = y_1$, it is called a \textbf{loop}. 
A loop that is homotopic to a single point, i.e.~a null-homotopic loop like the one in the example above, is also said to be \textbf{contractible} or \textbf{trivial}.  
A path-connected space in which every loop is contractible is called \textbf{simply connected}.  
So the real line and the surface of the sphere are simply connected, but the circle and the surface of a doughnut (the torus) are not.

\subsection{The fundamental group}

\begin{figure}
\centering
\begin{tabular}{ccc}
\includegraphics[width=0.3\textwidth]{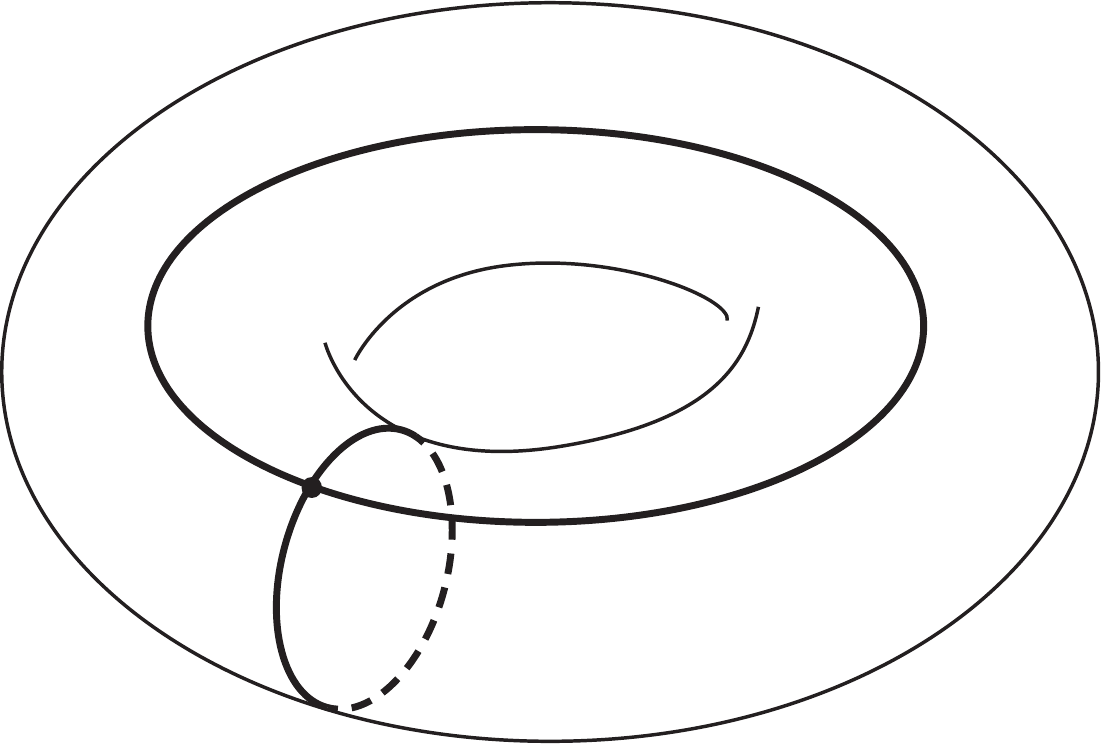} &
\includegraphics[width=0.3\textwidth]{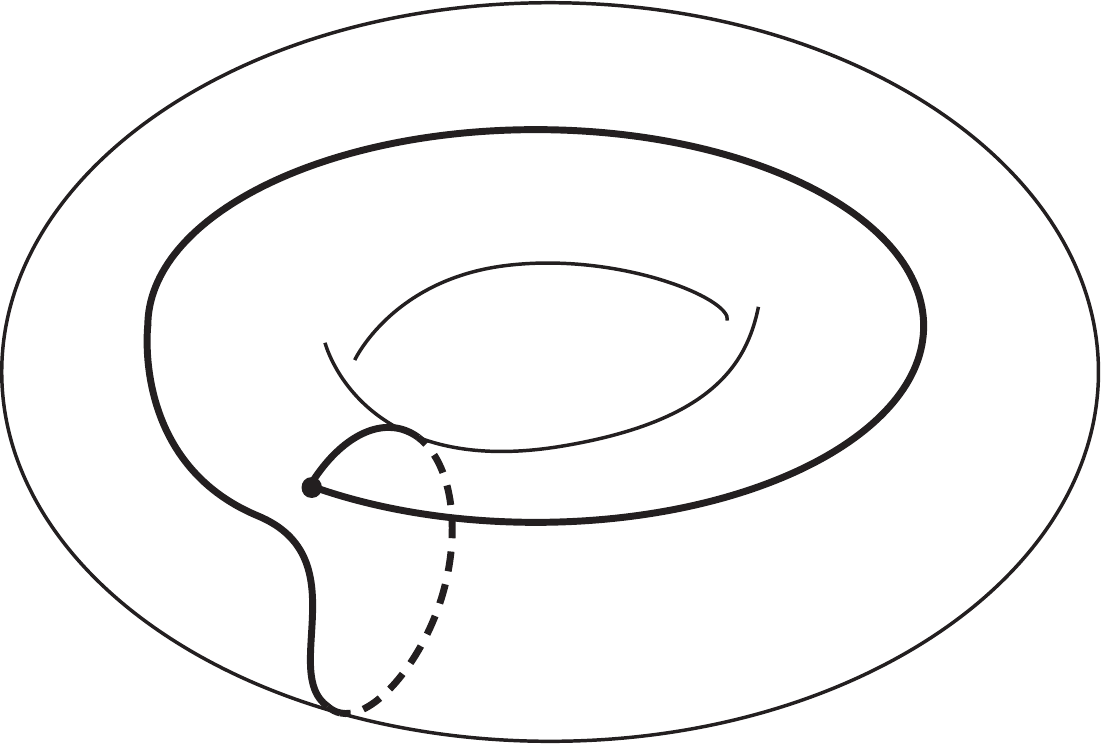} &
\includegraphics[width=0.3\textwidth]{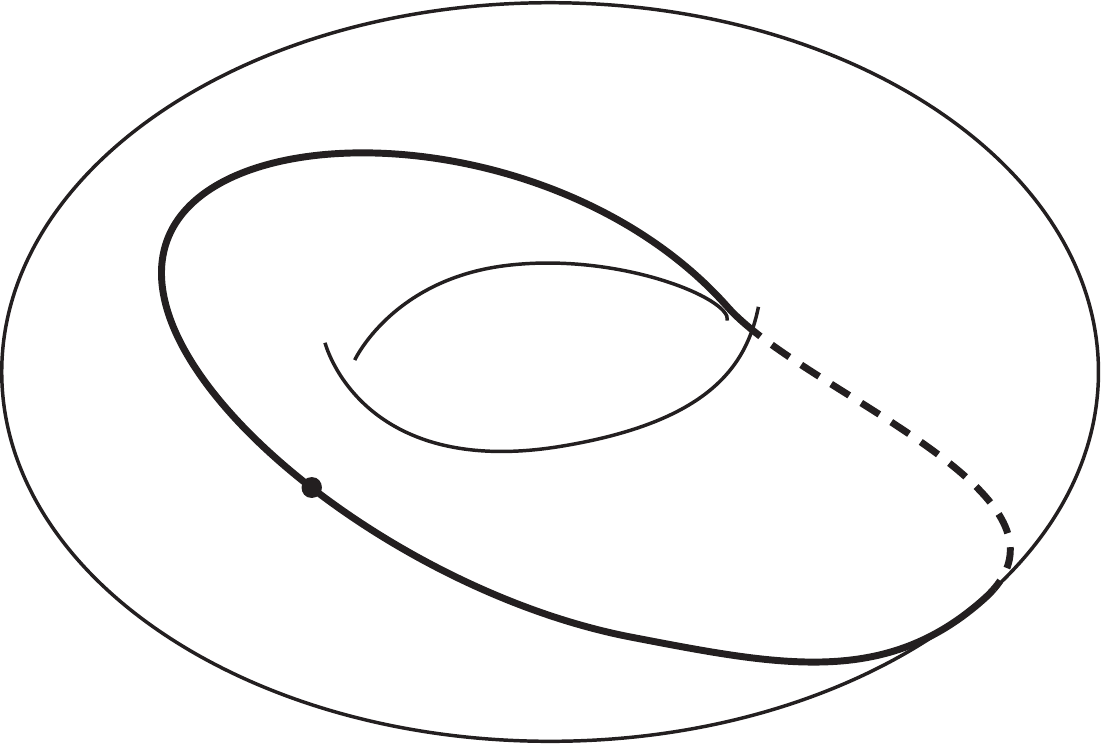} \\
a & b & c 
\end{tabular}
\caption{a) Two non-homotopic loops on the torus with the same base point. b) A loop homotopic to the concatenation the loops depicted in (a). c) Another loop in the same homotopy class.  }
\label{fig:torus}
\end{figure}

We are now in a position to define our first algebraic object, the \emph{fundamental group}.  
The first step is to choose a \textbf{base point} $y_0$ in the space $Y$ and consider all possible loops in $Y$ that start and end at $y_0$.  
Two loops belong to the same \textit{equivalence class} if they are homotopic: 
given a loop $f: [0,1] \to Y$ with $f(0) = f(1) = y_0$, we write $[f]$ to represent the set of all loops that are homotopic to $f$.  
The appropriate group operation $[f]*[g]$ on these equivalence classes is a concatenation of loops defined by tracing each at twice the speed. 
Specifically, choose $f$ and $g$ to be representatives of their respective equivalence classes and define $f*g(x) = f(2x)$ when $x \in [0,\tfrac{1}{2}]$  and $f*g(x) = g(2x-1)$ when $x \in [\tfrac{1}{2}, 1]$.  
Since all loops have the same base point, this product is another loop based at $y_0$. 
We then simply set $[f]*[g] = [f*g]$. 
The equivalence class of the product $[f*g]$ is independent of the choice of $f$ and $g$ because we can re-parameterize the homotopies in the same way as we concatenated the loops. Note, though, that the equivalence class $[f*g]$ consists of more than just concatenated loops; Fig.~\ref{fig:torus} depicts an example on the torus.       

The set of all homotopy equivalence classes of loops based at $y_0$ with the operation $*$ forms a group with the identity element being the class of null-homotopic loops $[e]$ where $e(x) = y_0$, and the inverse of a loop defined to be the same loop traced backwards:  $[f]^{-1} =  [ h ]$ where 
$h(x) = f(1-x)$.  
This group is the \textbf{fundamental group} of $Y$ with base point $y_0$:  $\pi_1( Y , y_0)$.  

The operation taking a topological space to its fundamental group is an example of a \textbf{functor}. 
This word expresses the property alluded to in the introduction that continuous maps between topological spaces transform to homomorphisms between their associated groups.   
The functorial nature of the fundamental group is manifest in the following fashion.  
Suppose we have a continuous function $f: X \to Y$ with $f(x_0) = y_0$.
Then given a loop in $X$ with base point $x_0$ we can use simple composition of the loop with $f$ to obtain a loop in $Y$ with base point $y_0$.  
Composition also respects the concatenation of loops and homotopy equivalences so it induces a homomorphism between the fundamental groups:  
$\pi_1(f) : \pi_1(X, x_0) \to \pi_1(Y,y_0)$.  
When the function $f:X \to Y$ is a \emph{homeomorphism}, it follows that the induced map $\pi_1(f)$ is an \emph{isomorphism} of their fundamental groups. 

Some further elementary properties of the fundamental group are: 
\begin{itemize}
\item A simply connected space has a trivial fundamental group, i.e.~only the identity element. 
\item If $Y$ is path-connected, the fundamental group is independent of the base point, and we write $\pi_1(Y)$. 
\item The fundamental group respects products\footnote{%
The (Cartesian or direct) \textbf{product} of two spaces (or two groups) $X \times Y$ is defined by ordered pairs $(x,y)$ where $x \in X$ and $y \in Y$.  
}
of path-connected spaces:  $\pi_1(X \times Y) =  \pi_1(X) \times \pi_1(Y)$. 
\item The wedge product of two path-connected spaces (obtained by gluing the spaces together at a single point) gives a free product\footnote{%
The \textbf{free product} of two groups $G*H$ is an infinite group that contains both $G$ and $H$ as subgroups and whose elements are words of the form $g_1 h_1 g_2 h_2 \cdots$.  
}
on their fundamental groups: $\pi_1(X \vee Y ) = \pi_1(X) * \pi_1(Y)$.   
\item The \textbf{van Kampen theorem} shows how to compute the fundamental group of a space 
$X = U \cup V$ when $U$, $V$, and $U \cap V$ are open, path-connected subspaces of $X$ via a \emph{free product with amalgamation}:  $\pi_1(X) = \pi_1(U) * \pi_1(V) / N$, where $N$ is a normal subgroup generated by elements of the form $ i_{U}(\gamma) i_{V}(\gamma)^{-1}$ and $\gamma$ is a loop in $\pi_1(U\cap V)$, $i_U$ and $i_V$ are inclusion-induced maps from $\pi_1(U\cap V)$ to $\pi_1(U)$ and $\pi_1(V)$ respectively.  See \cite{HatcherAT} for details. 
\end{itemize}

\subsection{Homotopy of spaces}
\label{HomotopyOfSpaces}

Now we look at what it means for two spaces $X$ and $Y$ to be \textbf{homotopy equivalent} or to have the same \textbf{homotopy type}: there must be continuous functions $f:X \to Y$ and $g:Y \to X$ such that $f  g : Y \to Y$ is homotopic to the identity on $Y$ and $g  f : X \to X$ is homotopic to the identity on $X$.
We can show that the unit circle $S^1$ and the annulus $A$ have the same homotopy type as follows.
Let 
\[
 S^1 = \{(r,\theta) \; | \; r = 1, \theta \in \left[0, 2\pi \right) \}  \quad \text{and} \quad  
 A = \{(r,\theta) \; | \;  1 \leq r \leq 2,  \theta \in \left[0, 2\pi \right)  \}
\]
be subsets of the plane parametrized by polar coordinates. 
Define $f : S^1 \to A$ to be the inclusion map $f(1,\theta) = (1,\theta)$ and let $g : A \to S^1$  map all points with the same angle to the corresponding point on the unit circle: $g(r,\theta) = (1,\theta)$. 
Then $gf: S^1 \to S^1$ is given by $gf(1,\theta) = (1,\theta)$ which is exactly the identity map.
The other composition is $fg : A \to A$ is $fg(r,\theta) = (1,\theta)$.
This is homotopic to the identity $i_A = (r,\theta)$ via the homotopy $F(r,\theta,t) = (1+t(r-1), \theta)$. 
This example is an illustration of a \textbf{deformation retraction}: a homotopy equivalence between a space (e.g.~the annulus) to a subset (the circle) that leaves the subset fixed throughout.  

Spaces that are homotopy equivalent have isomorphic fundamental groups.  
A space that has the homotopy type of a point is said to be \textbf{contractible} and has trivial fundamental group. 
This is much stronger than being simply-connected:  for example, the sphere $S^2$ is simply connected because every loop can be shrunk to a point, but it is not a contractible space.

\subsection{Examples} 
\label{Examples-Homotopy}

Real space $\mathbb{R}^m$, $m\geq1$, all spheres $S^n$ with $n\geq 2$, any Hilbert space, and any connected tree (cf. Chapter~\ref{GraphandNetworkTheory}) have trivial fundamental groups. 

The fundamental group of the circle is isomorphic to the integers under addition: $\pi_1(S^1) = \mathbb{Z}$. 
This can be seen by noting that the homotopy class of a loop is determined by how many times it wraps around the circle.  
A formal proof of this result is quite involved --- see Hatcher \cite{HatcherAT} for details.  
Any space that is homotopy equivalent to the circle will have the same fundamental group, this holds for the annulus, the M\"obius band, a cylinder, and the `punctured plane' $\mathbb{R}^2 \setminus (0,0)$.  

\begin{figure}
\centering
\includegraphics[width=0.3\textwidth]{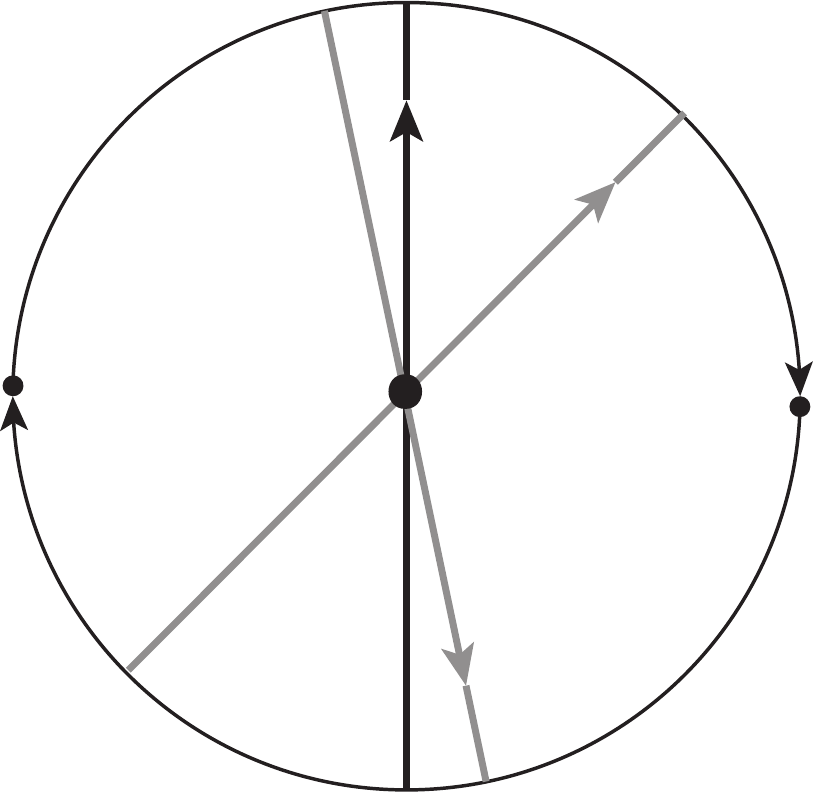}
\caption{The projective plane, $\mathbb{R}P^2$ is modelled by the unit disk with opposite points on the boundary identified.  The black loop starting at $(0,0)$ is homotopic to its inverse with the equivalence suggested by the grey loops. }
\label{fig:projective_plane}
\end{figure}

The projective plane $\mathbb{R}P^2$ is a non-orientable surface defined by identifying antipodal points on the boundary of the unit disk (or equivalently, antipodal points on the sphere). 
It has fundamental group isomorphic to $\mathbb{Z}_2$ (the group with two elements, the identity and $r$ which is its own inverse $r^2 =$ \textit{id}).  
To see this, consider a loop that starts at the center of the unit disk $(0,0)$, goes straight up to $(0,1)$ which is identified with  $(0,-1)$ then continues straight back up to the origin.  
This loop is in a distinct homotopy class to the null-homotopic loop but it is in the same homotopy class as its inverse (to see this, imagine fixing the loop at $(0,0)$ and rotate it by $180^{\circ}$ as illustrated in Fig.~\ref{fig:projective_plane}).  

The fundamental group of a connected graph with $v$ vertices and $e$ edges (cf. Chapter~\ref{GraphandNetworkTheory}) is a free group with $n$ generators\footnote{%
A \textbf{free group} with one generator, $a$ say, is the infinite cyclic group with elements $\ldots, a^{-1}, 1, a, a^2, \ldots$.  A free group with two generators $a, b$, contains all elements of the form $a^{i_1} b^{j_1} a^{i_2} b^{j_2} \cdots$ for integers $i_k$, $j_l$.   A free group with $n$ generators is the natural generalization of this. } 
where $n = e - (v - 1)$ is the number of edges in excess of a spanning tree.   This demonstrates that the fundamental group need not be Abelian (products do not necessarily commute).  

The torus $\mathbb{T} = S^1 \times S^1$ so $\pi_1(\mathbb{T}) = \mathbb{Z} \times \mathbb{Z}$.  More generally, an orientable genus-$g$ surface%
\footnote{Starting with a sphere, you can obtain all closed oriented 2-manifolds by attaching some number of handles (cylinders) to the sphere. The number of handles is the \textbf{genus}.} 
($g \geq 2$) has fundamental group isomorphic to a hyperbolic translation group with $2g$ generators. 

If a space has a finite \emph{cell structure}, then the fundamental group can be computed as a free group with relations in an algorithmic manner; this is discussed in Section~\ref{ComputationalTopology}. 


\subsection{Covering spaces}

The result about the fundamental group of a genus-$g$ surface comes from analyzing the relationship between loops on a surface and paths in its \emph{universal covering space} (the hyperbolic plane when $g\geq 2$). 
Covering spaces are useful in many other contexts (from harmonic analysis to differential topology to computer simulation), so we describe them briefly here.  
They are simply a more general formulation of the standard procedure of identifying a real-valued periodic function with a function on the circle. 

\begin{figure}
\centering
\includegraphics[width=0.4\textwidth]{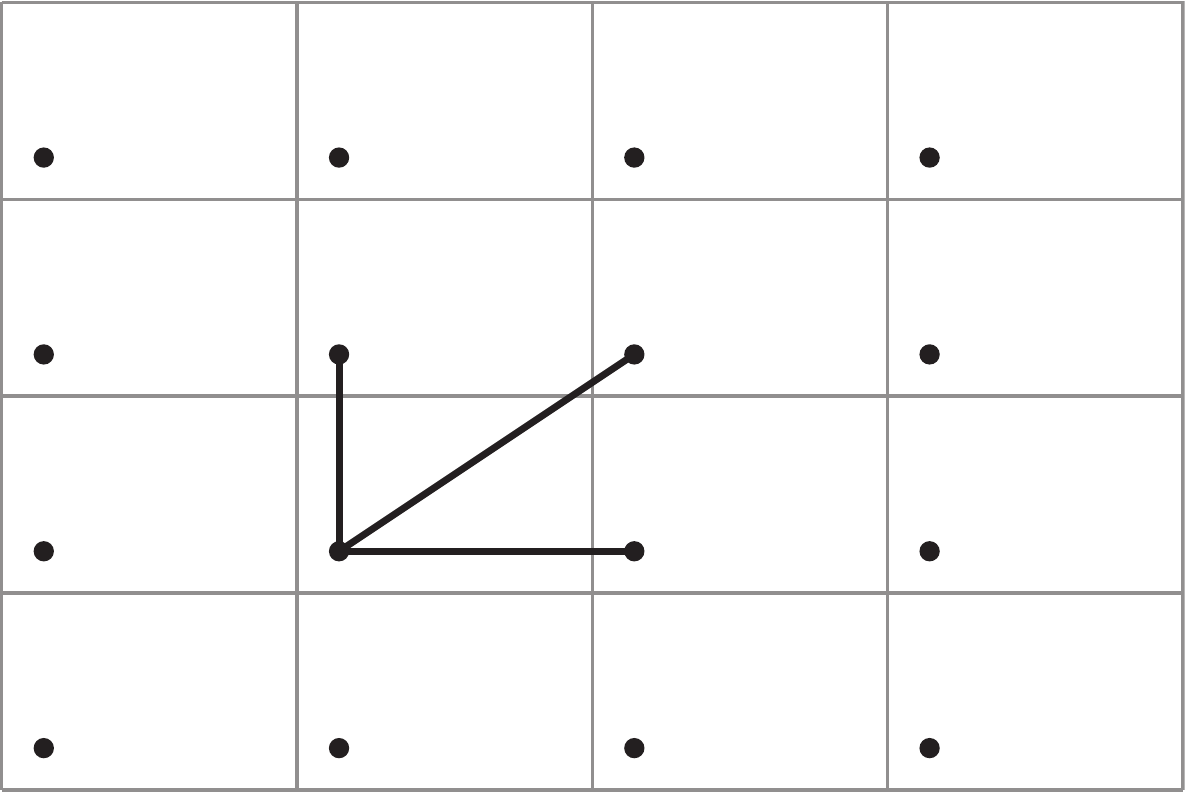}
\hspace{2mm}
\includegraphics[width=0.3\textwidth]{Figure2a_torus_generators.pdf}
\caption{The universal covering space of the torus is the Euclidean plane projected onto the closed surface by identifying opposite edges of each rectangle with parallel orientations.  The fibre of the base point on the torus is a lattice of points in the cover.  The two loops on the torus lift to the vertical and horizontal paths shown in the cover. The lift of the concatenation of these two loops (see Fig.~\ref{fig:torus} c) is homotopic to the diagonal path in the cover.  The deck transformation group for this cover is simply the group of translations that preserve the rectangles which is isomorphic to $\mathbb{Z} \times \mathbb{Z} = \pi_1(\mathbb{T})$.}
\label{fig:covering_space}
\end{figure}

Given a topological space $X$, a \textbf{covering space} for $X$ is a pair $(C, p)$, where $C$ is another topological space and $p: C \to X$ is a continuous function onto $X$.  
The \textbf{covering map} $p$ must satisfy the following condition: for every point $x \in X$, there is a neighborhood $U$ of $x$ such that $p^{-1}(U)$ is a disjoint union of open sets each of which is mapped homeomorphically onto $U$ by $p$.  
The discrete set of points $p^{-1}(x)$ is called the \textbf{fiber} of $x$.
A \textbf{universal} covering space is one in which $C$ is simply-connected.  
The reason for the name comes from the fact that a universal covering of $X$ will cover any other connected covering of $X$. 
For example, the circle is a covering space of itself with $C = S^1 = \{z \in \mathbb{C} : |z| = 1 \}$ and 
$p_k(z) = z^k $ for all non-zero integers $k$, while the universal cover of the circle is the real line with  $p_{\mathcal{U}} : \mathbb{R} \to S^{1}$ given by $p_{\mathcal{U}}(t) =  \exp(i 2\pi t)$.  
The point $z = (1,0) \in S^1$ is then covered by the fiber $t \in \mathbb{Z} \subset \mathbb{R}$. 
We illustrate a covering of the torus in Fig.~\ref{fig:covering_space}.  

When $X$ and $C$ are suitably nice spaces (connected and locally path connected), loops in $X$ based at $x_0$ \emph{lift} to paths in $C$ between elements of the fiber of $x_0$.  
So in the example of $S^1$, a path in $\mathbb{R}$ between two integers $i < j$ maps under $p_{\mathcal{U}}$ to a loop that wraps $j-i$ times around the circle.  

Now consider homeomorphisms of the covering space $h: C \to C$ that respect the covering map, $p(h(c)) = p(c)$.  
The set of all such homeomorphisms forms a group under composition called the \textbf{deck transformation group}.  
When $(C,p)$ is a universal covering space for $X$, it is possible to show that the deck transformation group must be isomorphic to the fundamental group of $X$. 
This gives a technique for determining the fundamental group of a space in some situations; see Hatcher~\cite{HatcherAT} for details and examples. 

\subsection{Extensions and applications}

As we saw in the examples of Section~\ref{Examples-Homotopy}, the fundamental group of an $n$-dimensional sphere is trivial for $n \geq 2$, so the question naturally arises how we might capture the different topological structures of $S^n$.  
To generalize the fundamental group, we examine maps from an 
$n$-dimensional unit cube $I^n$ into the space $X$ where the entire boundary of the cube is mapped to a fixed base point in $X$, i.e.  $f: I^n \to X$, with $f(\partial I^n) = x_0$.  
Elements of the \textbf{higher homotopy groups} $\pi_n(X, x_0)$ are then homotopy-equivalence classes of these maps.
The group operation is concatenation in the first coordinate just as we defined for one-dimensional closed paths above. 
The main difference in higher dimensions is that this operation now commutes.  

It is perhaps not too difficult to see that $\pi_2(S^2) = \mathbb{Z}$, although the multiple wrapping of the sphere by a piece of paper cannot be physically realized in $\mathbb{R}^3$ in the same way a piece of string wraps many times around a circle.  
The surprise comes with the result that $\pi_k(S^n)$ is non-trivial for most (but certainly not all) $k \geq n \geq 2$, and in fact mathematicians have not yet determined all the homotopy groups of spheres for arbitrary $k$ and $n$~\cite{HatcherAT}.  
Higher-order homotopy groups are a rich and fascinating set of topological invariants that are the  subject of active research in mathematics.

\begin{figure}
\centering
\includegraphics[width=0.4\textwidth]{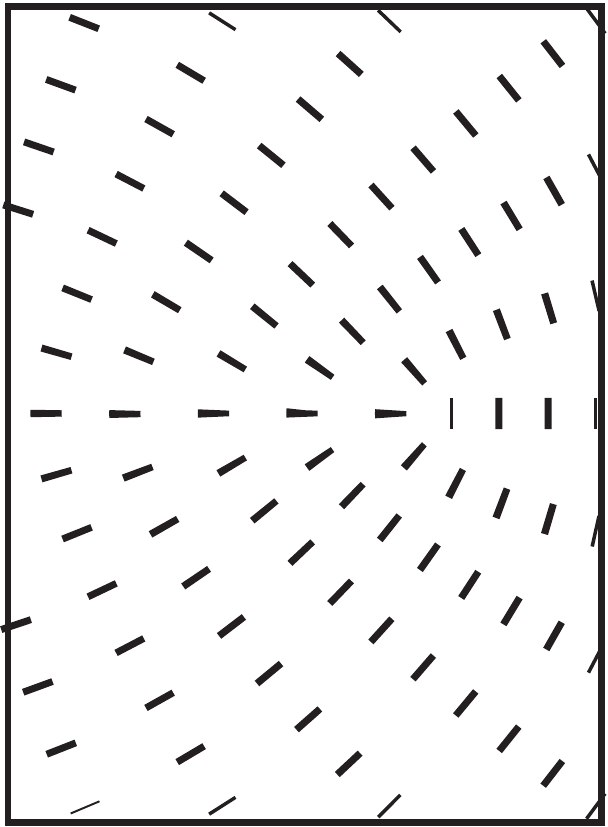}
\caption{A cross-section through a nematic fluid with a line defect that runs perpendicular to the page.  Each line-segment represents the averaged direction of a single molecule.  
}
\label{fig:nematic}
\end{figure}

One application of homotopy theory arises in the study of \textbf{topological defects} in condensed matter physics. 
A classic example is nematic liquid crystals which are fluids comprised of molecules with an elongated ellipsoidal shape.  
The order parameter for this system is the (time averaged) direction of the major axis of the ellipsoidal  molecule: $\mathbf{n}$. 
For identical and symmetrical molecules, the sign and the magnitude of the vector is irrelevant, and so the parameter space for $\mathbf{n}$ is the surface of the sphere with antipodal points identified --- topologically $\mathbb{R}P^2$.  
The existence of non-contractible loops in $\mathbb{R}P^2$ is associated with the existence of topological line defects in configurations of molecules in the nematic liquid crystal; see Fig.~\ref{fig:nematic}.  
The fact that $\pi_1(\mathbb{R}P^2) = \mathbb{Z}_2$ is manifest in the fact that two defects of the \emph{same} type can smoothly cancel one another.  
The second homotopy group $\pi_2(\mathbb{R}P^2) = \mathbb{Z}$, and this is manifest in the existence of point defects (``hedgehogs'') where the director field points radially away from a central point. 
See Mermin's original article \cite{Mermin79} or Nakahara \cite{Nakahara} for further details. 

\section{Homology}
\label{Homology}

The fundamental group is a useful invariant but it captures only the one-dimensional structure of equivalent loops in a space and cannot distinguish between spheres of dimensions greater than two, for example.  The higher homotopy groups do capture this structure but are difficult to compute.  
The homology groups provide a way to describe structure in all relevant dimensions, but require a bit more machinery to define.  This can seem abstract at first, but in fact the methods are quite combinatorial and there has been much recent activity devising efficient algorithms to compute  homological quantities from large data sets (see Section~\ref{ComputationalTopology}).

There are a number of different formulations of homology theory that give essentially the same results for `nice' spaces (such as differentiable manifolds).
The two key ingredients are a discrete cell complex that captures the way a space is put together, and a boundary map that describes incidences between cells of adjacent dimensions.  
The algebraic structure comes from defining the addition and subtraction of cells.  

The earliest formulation of homology theory is \textbf{simplicial homology}, based on triangulations of topological spaces called \textbf{simplicial complexes}. 
This theory has some short-comings when dealing with very general topological spaces and successive improvements over the past century have culminated in the current form based on singular homology and general cell complexes.    
Hatcher~\cite{HatcherAT} provides an excellent introduction to homology from this modern perspective.
We focus on simplicial homology here since it is the most concrete and easy to adapt for implementation on a computer. 
The notation used in this section is based on that of Munkres \cite{MunkresAT}.

\subsection{Simplicial complexes}

The basic building block is an \textbf{oriented $k$-simplex}, $\sigma^{k}$, the convex hull of $k+1$ geometrically independent points, $\{x_0,x_1,\ldots,x_k\} \subset \mathbb{R}^m$, with $k \leq m$.  
For example, a $0$-simplex is just a point, a $1$-simplex is a line segment, a $2$-simplex a triangle, and a $3$-simplex is a tetrahedron. 
We write $\sigma^k = \langle x_0,x_1,\ldots,x_k \rangle$ to denote a $k$-simplex and its vertices.  
The ordering of the vertices defines an \textbf{orientation} of the simplex.  
This orientation is chosen arbitrarily but is fixed, and coincides with the usual notion of orientation of line segments, triangles and tetrahedra.
Any even permutation of the vertices in a simplex gives another simplex with the same orientation, while an odd permutation gives a simplex with negative orientation.

Given a set $V$, an abstract \textbf{simplicial complex}, $\mathcal{C}$, is a collection of finite subsets of $V$ with the property if $\sigma^k = \{v_0, \ldots, v_k\} \in \mathcal{C} $ then all subsets of $\sigma^k$ (its faces) are also in $\mathcal{C}$. 
If the simplicial complex is finite then it can always be embedded in $\mathbb{R}^m$ for some $m$  (certain complexes with infinitely many simplices can also be embedded in finite-dimensional space).
An embedded complex is called a geometric realization of $\mathcal{C}$.
The subset of $\mathbb{R}^m$ occupied by the geometric complex is denoted by $|\mathcal{C}|$ and is called a \textbf{polytope} or \textbf{polyhedron}. 
When a topological space $X$ is homeomorphic to a polytope, $|\mathcal{C}|$, it is called a \textbf{triangulated space} and the simplicial complex $\mathcal{C}$ is a \textbf{triangulation} of $X$.  
For example, a circle is homeomorphic to the boundary of a triangle so the three vertices $a,b,c$ and three $1$-simplices, $\langle ab \rangle, \langle bc \rangle, \langle ca \rangle$ are a triangulation of the circle (see Fig.~\ref{fig:triangle}).  
All \emph{differentiable} manifolds have triangulations, but  a complete characterization of the class of topological spaces that have a triangulation is not known.  
Every topological 2 or 3-manifold has a  triangulation, but there is a (non-smooth) 4-manifold that cannot have a triangulation (it is related to the Lie group $E_8$ \cite{Scorpan}). 
The situation for non-differentiable manifolds in higher dimensions remains uncertain.

\subsection{Simplicial homology groups}

We now define the group structures associated with a space $X$ that is triangulated by a finite simplicial complex $\mathcal{C}$.
Although the triangulation of a space is not unique, the homology groups for any triangulation of the same space are identical (see \cite{MunkresAT}); this makes simplicial homology well-defined.   

The set of all $k$-simplices from $\mathcal{C}$ form the basis of a free group called the $k$-th \textbf{chain group}, $C_k(X)$. 
The group operation is an additive one; recall that $-\sigma^k$ is just $\sigma^k$ with the opposite orientation, so this defines the inverse elements.    
In general, a $k$-chain is the formal sum of a finite number of oriented $k$-simplices: $c_k = \sum_i a_i \sigma^k_i$.
The coefficients, $a_i$, are elements of another group, called the \textbf{coefficient group} that is typically the integers, $\mathbb{Z}$, but can be any Abelian group such as the integers mod 2 $\mathbb{Z}_2$, the rational $\mathbb{Q}$, or real numbers $\mathbb{R}$. 
If the coefficient group $G$ needs to be emphasized we write $C_k(X;G)$.   

When $k\geq1$ the \textbf{boundary operator} $\partial_{k} : C_{k} \to C_{k-1}$ maps a $k$-simplex onto the sum of the $(k-1)$-simplices in its boundary. 
If $\sigma^k =  \langle x_0,x_1,\ldots,x_k \rangle$ is a $k$-simplex, we have 
\begin{equation*}
 \partial_k (\sigma^k) = \sum_{i=0}^{k} (-1)^i \langle x_0,\ldots,\hat{x}_i,\ldots,x_k \rangle
\end{equation*}
where $ \langle x_0,\ldots,\hat{x}_i,\ldots,x_k \rangle $ represents the $(k-1)$-simplex obtained by deleting the vertex $x_i$.
The action of the boundary operator on general $k$-chains is obtained by linear extension from its action on the $k$-simplices:
$ \partial_k (\sum_i a_i \sigma^k_i) =  \sum_i a_i \partial_k ( \sigma^k_i)$.
For $k = 0$ the boundary operator is defined to be null:  $\partial_0 (c_0) = 0$.  
We drop the subscript from the boundary operator when the dimension is understood.

\begin{figure}
\centering
\includegraphics[width=0.3\textwidth]{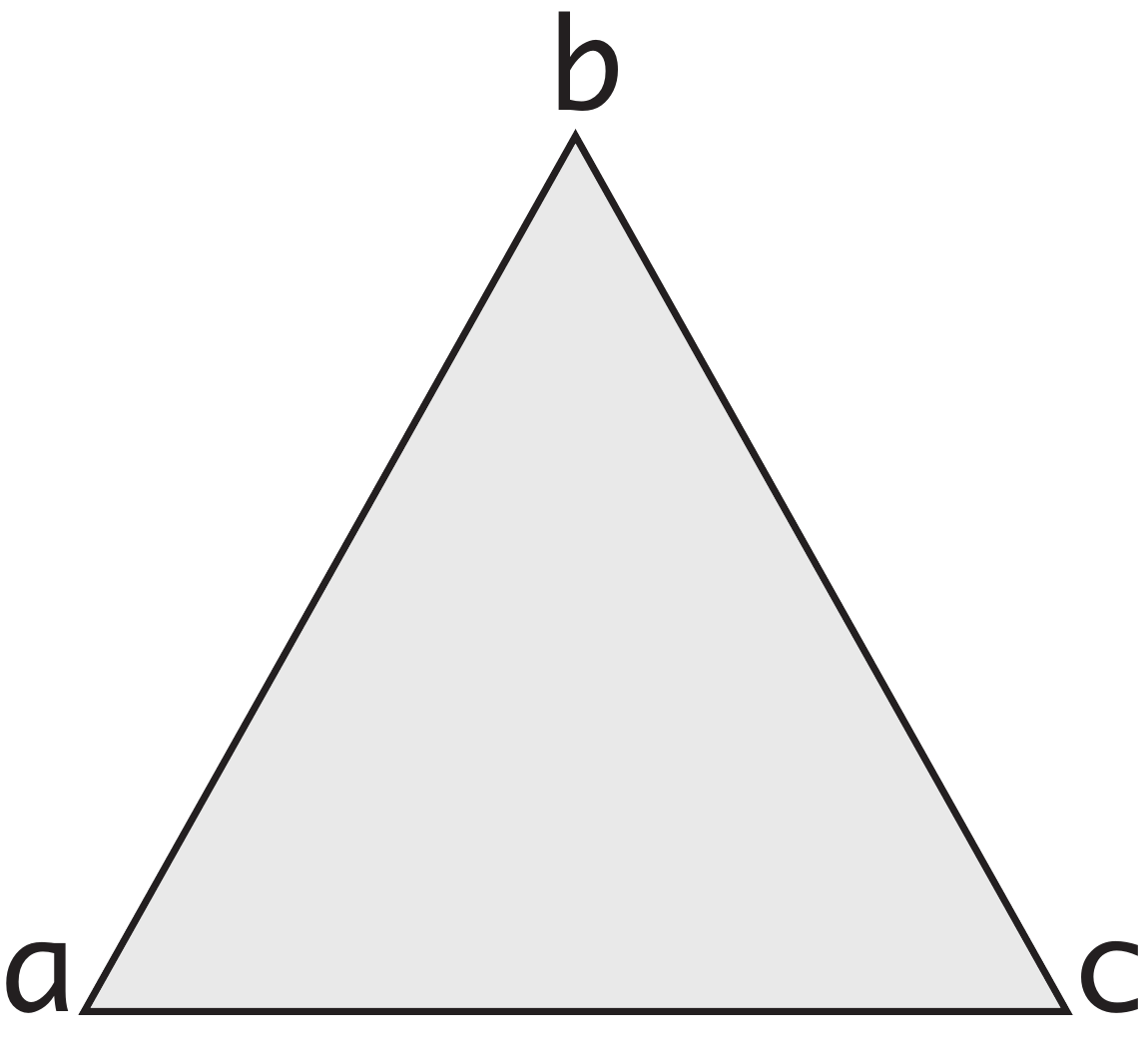}
\caption{The simplicial complex of a triangle consists of one 2-simplex, three 1-simplices and three vertices (0-simplices).}
\label{fig:triangle}
\end{figure}

As an example, consider the simplicial complex consisting of a triangle and all its edges and vertices, as shown in Fig.~\ref{fig:triangle}. 
The boundary of the $2$-simplex $\langle a,b,c \rangle$  is 
\begin{equation*} 
\partial (\langle a,b,c \rangle) = \langle b,c \rangle - \langle a,c \rangle + \langle a,b \rangle, 
\end{equation*}
and the boundary of this $1$-chain is:
\begin{equation*}
 \partial ( \langle b,c \rangle - \langle a,c \rangle + \langle a,b \rangle ) = (c - b) - (c - a) + (b - a) = 0.
\end{equation*}
This illustrates the fundamental property of the boundary operator, namely that
\begin{equation} 
\label{eq:bdybdy=0}
 \partial_{k} \, \partial_{k+1} = 0 
\end{equation}

We now consider two subgroups of $C_k$ that have important geometric interpretations. 
The first subgroup consists of $k$-chains that map to zero under the boundary operator. 
This group is the \textbf{group of cycles} denoted $Z_k$, it is the kernel (or null space) of $\partial_k$ and its elements are called $k$-cycles.   
From the definition of $\partial_0$ we see that all 0-chains are cycles so $Z_0 = C_0$.
The second subgroup of $C_k$ is the group of $k$-chains that bound a $(k+1)$-chain. 
This is the \textbf{group of boundaries} $B_k$, it is the image of $\partial_{k+1}$. 
It follows from (\ref{eq:bdybdy=0}) that every boundary is a cycle,
i.e. the image of $\partial_{k+1}$ is mapped to zero by $\partial_k$ so $B_k$ is a subgroup of $Z_k$.  
In our example of the triangle simplicial complex, we find no 2-cycles and a 1-cycle, 
$\langle b,c \rangle - \langle a,c \rangle + \langle a,b \rangle$, such that all other 1-cycles are integer multiples of this.  
This 1-cycle is also the only 1-boundary so $Z_1 = B_1$, and the 0-boundaries $B_0$ are generated by the two 0-chains:  $c-b$ and $c-a$ (the third boundary $a-b = (c-b) - (c-a)$).   

Since $B_k \subset Z_k$, we can form a quotient group, $H_k = Z_k/B_k$; this is the precisely the $k$-th \textbf{homology group}. 
The elements of $H_k$ are equivalence classes of $k$-cycles that do not bound any $k+1$ chain so  this is how homology characterizes $k$-dimensional holes.  
Formally, two $k$-cycles $w, z \in Z_k$ are in the same equivalence class if $w-z \in B_k$; such cycles are said to be homologous. 
We write $[z] \in H_k$ for the equivalence class of cycles homologous to $z$.
For the simple example of the triangle simplicial complex, we have already seen that 
$Z_1 = B_1$ so that $H_1 = \{ 0 \}$. 
The 0-cycles are generated by $\{ a, b, c \}$ and the  boundaries by $\{(c-b), (c-a)\}$, so $H_0$ has a single equivalence class, $[c]$, and is isomorphic to $\mathbb{Z}$.

The homology groups for some familiar spaces are: 
\begin{itemize}
\item  Real space $\mathbb{R}^n$  has $H_0 = \mathbb{Z}$ and $H_k = \{ 0 \}$ for $k \geq 1$.  
\item The spheres have $H_0(S^n) = \mathbb{Z}$,  $H_n(S^n) = \mathbb{Z}$ and 
$H_k(S^n) = \{0 \}$ for all other values of $k$. 
\item The torus has $H_0 = \mathbb{Z}$,  $H_1 = \mathbb{Z} \oplus \mathbb{Z}$, $H_2 = \mathbb{Z}$, $H_k = \{0\}$ for all other $k$.  The 2-cycle that generates $H_2$ is the entire surface.  This is in contrast to the second homotopy group for the torus, which is trivial.   
\item The real projective plane, $\mathbb{R}P^2$ has $H_0 = \mathbb{Z}$, $H_1 = \mathbb{Z}_2$,  and $H_k = \{0 \}$ for $k\geq 2$. The fact that $H_2$ is trivial is a result of the surface being \emph{non-orientable}; even though $\mathbb{R}P^2$ is \emph{closed} as a manifold, the 2-chain covering the surface is not a 2-cycle.  
\end{itemize}
The combinatorial nature of simplicial homology makes it readily computable.  We give the classical algorithm and review recent work on fast and efficient algorithms for data in Section~\ref{ComputationalTopology}.

\subsection{Basic properties of homology groups} 

In general, the homology groups of a finite simplicial complex are finitely generated Abelian groups, so the following theorem tells us about their general structure (see Theorem 4.3 of \cite{MunkresAT}). 
\begin{theorem}
\label{thm:abgrp}
If G is a finitely generated Abelian group then it is isomorphic to the following direct sum:
\begin{equation*} \label{eq:abgrp}
  G \simeq (\mathbb{Z} \oplus \cdots \oplus \mathbb{Z}) \oplus 
              \mathbb{Z} _{t_1} \oplus \cdots \oplus \mathbb{Z}_{t_m} . 
\end{equation*}
\end{theorem}
The number of copies of the integer group $\mathbb{Z}$ is called the \textbf{Betti number} $\beta$.  
The cyclic groups $\mathbb{Z}_{t_i}$ are called the \emph{torsion subgroups} and the $t_i$ are the \textbf{torsion coefficients}; they are defined so that $t_i > 1$ and $t_1$ divides $t_2$ which divides $t_3$ and so on. 
The torsion coefficients of the homology group $H_k(\mathcal{C})$ measure the twistedness of the space in some sense.  For example, the real projective plane has  $H_1 = \mathbb{Z}_2$, because the 2-chain that represents the whole of the surface has a boundary that is twice the generating 1-cycle.  
The Betti number $\beta_k$ of the $k$-th homology group counts the number non-equivalent non-bounding $k$-cycles and this can be loosely interpreted as the number of $k$-dimensional holes.  
The 0-th Betti number, $\beta_0$, counts the number of path-connected components of the space.

Some other fundamental properties of the homology groups are as follows:
\begin{itemize}
\item If the simplicial complex has $N$ connected components, $X = X_1 \cup \cdots \cup X_N$ then $H_0$ is isomorphic to the direct sum of $N$ copies of the coefficient group, and 
$H_k(X) = H_k(X_1) \oplus \cdots \oplus H_k(X_N)$. 
\item Homology is a functor.  If $f: X \to Y$ is a continuous function from one simplicial complex into another, it induces natural maps on the chain groups $f_{\sharp}: C_k(X) \to C_k(Y)$ for each $k$ which commute with the boundary operator:  $\partial f_{\sharp} = f_{\sharp} \partial$.   This commutativity implies that cycles map to cycles and boundaries to boundaries, so that the $f_{\sharp}$ induce homomorphisms on the homology groups $f_{*} : H_k(X) \to H_k(Y)$.  
\item If two spaces are homotopy equivalent they have isomorphic homology groups (this is shown using the above functorial property). 
\item The first homology group is the \emph{Abelianization} of the fundamental group.  
When $X$ is a path-connected space, the connection between $H_1(X)$ and $\pi_1(X)$ is made by noticing that two 1-cycles are equivalent in homology if their difference is the boundary of a 2-chain; if we parametrize the 1-cycles as loops then this 2-chain forms a region through which one can define a homotopy. See \cite{HatcherAT} for a formal proof. 
\item The higher-dimensional homology groups have the comforting property that if all simplices in a complex have dimensions $\leq m$ then $H_k = \{0 \}$ for $k > m$.  This is in stark contrast to the higher-dimensional homotopy groups.  
\end{itemize} 

A particularly pleasing result in homology relates the Betti numbers to another topological invariant called the \textbf{Euler characteristic}.  \label{eulerchar}
For a finite simplicial complex, $\mathcal{C}$, define $n_k$ to be the number of simplices of dimension $k$, then the Euler characteristic is defined to be 
$ \chi ( \mathcal{C} ) =  n_0 - n_1 + n_2 - \cdots $.  
The \emph{Euler-Poincar\'e theorem} states that the the alternating sum of Betti numbers is the same as the Euler characteristic~\cite{MunkresAT}:   $\chi = \beta_0 - \beta_1 + \beta_2 - \cdots$.   
This is one of many results that connect the Euler characteristic with other properties of manifolds.  
For example, if $M$ is a compact 2-manifold with a Riemannian metric, then the \emph{Gauss-Bonnet theorem} states that $2\pi \chi$ is equal to the integral of Gaussian curvature over the surface plus the integral of geodesic curvature over the boundary of $M$ \cite{Hyde_shape_book}.   
Further, if $M$ is orientable and has no boundary then it must be homeomorphic to a sphere with $g$ handles and $\chi = 2 - 2g$ where $g$ is the genus of the surface.  
When $M$ is non-orientable without boundary, then it is homeomorphic to a sphere with $r$ cross-caps and $\chi = 2-r$.   

The Euler characteristic is a topological invariant with the property of \emph{inclusion-exclusion}: 
If a triangulated space $X = A \cup B$ where $A$ and $B$ are both subcomplexes, then 
\begin{equation*}
\chi(X) = \chi(A) + \chi(B) - \chi(A \cap B).  
\end{equation*}
This means the value of $\chi$ is a localizable one and can be computed by cutting up a larger space into smaller chunks.
This property makes it a popular topological invariant to use in applications~\cite{Mecke98}.   
A recent application that exploits the local additivity of the Euler characteristic to great effect is target enumeration in localized sensor networks~\cite{Ghrist09,Ghrist12}.   
The Euler characteristic has also been shown to be an important parameter in the physics of porous materials~\cite{Arns03,Scholz2012}.  

The simple inclusion-exclusion property above does not hold for the Betti numbers since they capture global aspects of the topology of a space. 
Relating the homology of two spaces to their union requires more sophisticated algebraic machinery that we review below.

\subsection{Homological algebra}

Many results and tools in homology are independent of the details about the way the chains  and boundary operators are defined for a topological space; they depend only on the chain groups and the fact that $\partial \partial = 0$.   
The study of such abstract chain complexes and transformations between them is called \textbf{homological algebra} and is one of the original examples in \emph{category theory}~\cite{MunkresAT}.    

An abstract \textbf{chain complex} is a sequence of Abelian groups and homomorphisms  
\begin{equation*}
 \cdots \stackrel{d_{k+2}}{\longrightarrow} C_{k+1}  \stackrel{d_{k+1}}{\longrightarrow}  C_{k} \stackrel{d_{k}}{\longrightarrow}  \cdots \stackrel{d_1}{\longrightarrow}  C_0 \longrightarrow \{0\} ,  \quad  \text{with } d_{k}d_{k+1} = 0.   
\end{equation*}
The homology of this chain complex is  $H_k (C) = \text{ker } d_k / \text{im } d_{k+1}$.
In certain cases (such as the simplicial chain complex of a contractible space) we find that 
$\text{im } d_{k+1} = \text{ker } d_k$ for $k\geq1$ so the homology groups are trivial. 
Such a sequence is said to be \textbf{exact}.  

This property of exactness has many nice consequences.  
For example, with a \textbf{short exact sequence} of groups 
\begin{equation*}
0 \to A \stackrel{f}{\longrightarrow} B \stackrel{g}{\longrightarrow} C \to 0.
\end{equation*}
the exactness means that $f$ is a monomorphism (one-to-one) and $g$ is an epimorphism (onto). In fact $g$ induces an isomorphism of groups $C \approx B/ f(A)$,  and if these groups are finitely generated Abelian then the Betti numbers satisfy $\beta(B) =  \beta(C) + \beta(A)$ (where we have replaced $f(A)$ by $A$ since $f$ is one-to-one).    

Now imagine there is a short exact sequence of \emph{chain complexes}, i.e. 
\begin{equation*}
0 \to A_k \stackrel{f_k}{\longrightarrow} B_k \stackrel{g_k}{\longrightarrow} C_k \to 0.
\end{equation*}
is exact for all $k$ and the maps $f_k$, $g_k$ commute with the boundary operators in each complex (i.e. $d_B f_k = f_{k-1} d_A$, etc.). 
Typically, the $f_k$ will be induced by an inclusion map (and so be monomorphisms), and $g_k$ by a quotient map (making them epimorphisms) on some underlying topological spaces.  
The \textbf{zig-zag lemma} shows that these short exact sequences can be joined together into a long exact sequence on the homology groups of $A, B$ and $C$: 
\begin{equation*}
  \cdots \to H_k(A)  \stackrel{f_{*}}{\longrightarrow} H_k(B) \stackrel{g_{*}}{\longrightarrow}  H_k(C) 
  \stackrel{\Delta}{\longrightarrow}  H_{k-1}(A)  \stackrel{f_{*}}{\longrightarrow} H_{k-1}(B) \to \cdots 
\end{equation*}
The maps $f_{*}$ and $g_{*}$ are those induced by the chain maps $f$ and $g$ and the boundary maps $\Delta$ are defined directly on the homology classes in $H_k(C)$ and $H_{k-1}(A)$ by showing that cycles in $C_k$ map to cycles in $A_{k-1}$ via $g_k$, the boundary $\partial_B$, and $f_{k-1}$.   Details are given in Hatcher \cite{HatcherAT}. 


One of the most useful applications of this long exact sequence in homology is the \textbf{Mayer-Vietoris exact sequence}, a result that describes the relationship between the homology groups of two simplicial complexes, $X, Y$, their union and their intersection.  This gives us a way to deduce homology groups of a larger space from smaller spaces.  
\begin{multline}\label{Mayer-Vietoris}
\cdots  \stackrel{j_{k}}{\longrightarrow} H_{k}(X) \oplus  H_{k}(Y)  
   	   \stackrel{s_{k}}{\longrightarrow}  H_{k}(X \cup Y)  \stackrel{v_{k}}{\longrightarrow}  \\
	   H_{k-1}(X \cap Y ) 
	   \stackrel{j_{k-1}}{\longrightarrow}  H_{k-1}(X) \oplus H_{k-1}(Y) 
	   \stackrel{s_{k-1}}{\longrightarrow}  \cdots 
\end{multline}
The homomorphisms are defined as follows: 
\begin{align*}
j_k([u]) &= ( [u], - [u] ) \\
s_k([w], [w']) &= [w+ w'] \\
v_k([z]) &= [\partial z'],
\end{align*}
where in the last equation $z$ is a cycle in $X\cup Y$ and we can write $z = z' + z''$ where $z'$ and  $z''$ are chains (not necessarily cycles) in  $X$ and $Y$ respectively. 
These homomorphisms are well defined (see, for example, Theorem 33.1 of \cite{MunkresAT}). 
Exactness implies that the image of each homomorphism is equal to the kernel of the following one:  $\text{im } j_p = \text{ker } s_p$,  $\text{im } s_p = \text{ker } v_p$, and $\text{im } v_p = \text{ker } j_{p-1}$. 

The Mayer-Vietoris sequence has an interesting interpretation in terms of Betti numbers.  
First we define $N_k = \text{ker }j_k$ to be the subgroup of $H_k(X \cap Y)$ defined by the $k$-cycles that bound in both $X$ and in $Y$.  
Then by focussing on the exact sequence around $H_k(X\cup Y)$ it follows that \cite{Delfinado93,AlexandroffHopf} 
\begin{equation*}
  \beta_k(X\cup Y) = \beta_k(X) + \beta_k(Y) - \beta_k(X\cap Y) + \text{rank }N_k + \text{rank }N_{k-1}. 
\end{equation*}
This is where we see the non-localizable property of homology and the Betti numbers most clearly.  

The Mayer-Vietoris sequence holds in a more general setting than simplicial homology.  It is an example of a result that can be derived from the Eilenberg-Steenrod axioms.   Any theory for which these five axioms hold is a type of homology theory, see \cite{HatcherAT} for further details.

\subsection{Other homology theories} 

Chain complexes that capture topological information may be defined in a number of ways.  We have defined simplicial chain complexes above, and will briefly describe some other techniques here. 

\textbf{Cubical homology} is directly analogous to simplicial homology using $k$-dimensional cubes as building elements rather than $k$-dimensional simplices.  This theory is developed in full in \cite{kaczynskibook} and arose from applications in digital image analysis and numerical  analysis of dynamical systems.  

\textbf{Singular homology} is built from singular $k$-simplices which are simply continuous functions from the standard $k$-simplex into a topological space $X$, $\sigma:  \langle v_0,\ldots,v_k \rangle \to X$.  
Singular chains and the boundary operator are defined as they are in simplicial homology.  
A greater degree of flexibility is found in singular homology since the maps $\sigma$ are allowed to collapse the simplices, e.g. the standard $k$-simplex, $k>0$ may have boundary points mapping to the same point in $X$, or the entire simplex may be mapped to a single point \cite{HatcherAT}. 

An even more general formulation of \textbf{cellular homology} is made by considering general cell (CW) complexes.  A cell complex is built incrementally by starting with a collection of points $X^{(0)} \subset X$, then  attaching $1$-cells via maps of the unit interval into $X$ so that end points map into $X^{(0)}$ to form the 1-skeleton $X^{(1)}$.  This process continues by attaching $k$-cells to the $(k-1)$-skeleton by continuous maps of the closed unit $k$-ball, $\phi: B_{k} \to X$ that are homeomorphic on the interior and satisfy $\phi: \partial B_{k} \to X^{(k-1)}$.  
The definition of the boundary operator for a cell complex requires the concept of \emph{degree} of a map of the $k$-sphere (i.e. the boundary of a $(k+1)$-dimensional ball). 
For details see Hatcher \cite{HatcherAT}. 

We will see in the section on Morse Theory that it is also possible to define a chain complex from the  critical points of a smooth function on a manifold.

\section{Cohomology} 
\label{Cohomology}

The cohomology groups are derived by a simple dualization procedure on the chain groups (similar to the construction of dual function spaces in analysis).   
We will again give definitions in the simplicial setting but the concepts carry over to other contexts.  
A \textbf{cochain} $\phi^k$ is a function from the simplicial chain group into the coefficient group,  $\phi^k: C_k (X;G) \to G$ (recall that $G$ is usually the integers, $\mathbb{Z}$, but can be any Abelian group).  
The space of all $k$-cochains forms a group called the $k$-th \textbf{cochain group} $C^k(X;G)$.  
The simplicial boundary operators $\partial_k : C_k \to C_{k-1}$ induce \textbf{coboundary} operators $\delta^{k-1} : C^{k-1} \to C^{k}$ on the cochain groups via $\delta (\phi) = \phi \partial$.   
In other words, the cochain $\delta (\phi)$ is defined via the action of $\phi$ on the boundary of each $k$-simplex $\sigma = \langle x_0, x_1, \ldots, x_k \rangle$: 
\begin{equation*}
  \delta(\phi) (\sigma) = \sum_{i} (-1)^i  \phi ( \langle x_0,\ldots,\hat{x}_i,\ldots,x_k \rangle ).  
\end{equation*}
The key property from homology that $\partial_k \partial_{k+1} = 0$ also holds true for the coboundary: $\delta^{k} \delta^{k-1}= 0$ (coboundaries are mapped to zero) so we define the $k$-th \textbf{cohomology group} as $H^k (X) = \text{ker } \delta^k / \text{im } \delta^{k-1}$.  
Note that cochains $\phi \in \text{ker } \delta$ are functions that vanish on the $k$-boundaries (not the larger group of $k$-cycles), and a coboundary $\eta^{k} \in \text{im } \delta$ is one that can be defined via the action of some cochain $\phi^{k-1}$ on the $(k-1)$-boundaries. 

The coboundary operator acts in the direction of increasing dimension and this can be a more natural action in some situations (such as de Rham cohomology of differential forms discussed below) and also has some interesting algebraic consequences (it leads to the definition of the cup product). 
\begin{equation*}
 \cdots {\longleftarrow} C^{k+1}  \stackrel{\delta^{k}}{\longleftarrow}  C^{k} \longleftarrow  \cdots \longleftarrow  C^0 \longleftarrow \{ 0 \} . 
\end{equation*}
This action of the coboundary makes cohomology \textbf{contravariant} (induced maps act in the opposite direction) where  homology is \textbf{covariant} (induced maps act in the same direction).
If $f : X\to Y$ is a continuous function between two topological spaces then the group homomorphism induced on the cohomology groups acts as $f^* : H^k(Y) \to H^k(X)$. 

In simplicial homology, the simplices form bases for the chain groups, and we can similarly use them as bases for the cochain groups by defining an \textbf{elementary cochain} $\dot{\sigma}$ as the function that takes the value one on $\sigma$ and zero on all other simplices.  
For a finite simplicial complex it is possible to represent the boundary operator $\partial$ as a matrix with respect to the bases of oriented $k$- and $(k-1)$-simplices.    
If we then use the elementary cochains as bases for the cochain groups, the matrix representation for the coboundary operator is just the transpose of that for the boundary operator.  
This shows that for finite simplicial complexes, the functional and geometric meanings of duality are the same.   

Another type of duality is that between homology and cohomology groups on compact closed oriented manifolds (i.e.~without boundary).  
\textbf{Poincar\'e duality} states that $H^k (M) = H_{m-k} (M)$ for $k \in \{0,\ldots,m\}$ where $m$ is the dimension of the manifold, $M$; see \cite{HatcherAT} for further details.  

Despite this close relationship between homology and cohomology on manifolds, the cohomology groups have a naturally defined product combining two cochains and this additional structure can help distinguish between some spaces that homology does not.   
We start with $\phi \in C^k(X;G)$ and $\psi \in C^l (X;G)$ where the coefficient group should now be a ring $R$ (i.e.~$R$ should have both addition and multiplication operations; $\mathbb{Z}$, $\mathbb{Z}_p$, and $\mathbb{Q}$ are rings.) 
The \textbf{cup product} is the cochain $\phi \smallsmile \psi \in C^{k+l}(X;R)$ defined by its action on a $(k+l)$-simplex $\sigma = \langle v_0, \ldots, v_{k+l} \rangle$ as follows:
\begin{equation*}
	(\phi \smallsmile \psi )(\sigma) = \phi( \langle v_0, \ldots, v_k \rangle)  \psi (\langle v_k, \ldots, v_{k+l} \rangle) .  
\end{equation*}
The relation between this product and the coboundary is:
\begin{equation*}
         \delta (\phi \smallsmile \psi)  =  \delta \phi \smallsmile \psi + (-1)^{k} \phi \smallsmile \delta \psi . 
\end{equation*}
From this, we see that the product of two cocycles is another cocycle, and if the product is between a cocycle and a coboundary, then the result is a coboundary. 
Thus, the cup product is a well defined product on the cohomology groups that is \emph{anticommutative}:  $[\phi] \smallsmile [\psi] = (-1)^{kl} [\psi] \smallsmile [\phi]$ (provided the coefficient ring, G,  is commutative). 
These rules for products of cocycles should look suspiciously familiar to those who have read Chapter~\ref{DifferentialGeometry}.  They are similar to those for exterior products of differential forms  and this relationship is formalized in the next section when we define de Rham cohomology.

\subsection{De Rham cohomology}

One interpretation of cohomology that is of particular interest in physics comes from the study of \emph{differential forms} on smooth manifolds; cf.~Chapter~\ref{DifferentialGeometry}.  
Recall that a differential form of degree $k$, $\omega$, defines for each point $p \in M$, an alternating multilinear map on $k$ copies of the tangent space to $M$ at $p$:  
\begin{equation*}
    \omega_p  :  T_p M \times \cdots \times T_p M  \to  \mathbb{R} 
\end{equation*} 
The set of all differential $k$-forms on a manifold $M$ is a vector space, $\Omega^k (M)$, 
and the \textbf{exterior derivative} is a linear operator that takes a $k$-form to a $k+1$-form, 
$ d_k :  \Omega^k(M) \to \Omega^{k+1}(M)$ as defined in Chapter~\ref{DifferentialGeometry}. 

The crucial property $dd = 0$ holds for the exterior derivative.  In this context, $k$-forms in the image of $d$ are called $\textbf{exact}$, i.e. $\omega = d \sigma$ for some $(k-1)$-form $\sigma$; and those for which $d \omega = 0$ are called $\textbf{closed}$.   
We therefore have a cochain complex of differential forms and can form quotient groups of closed forms modulo the exact forms to obtain the de Rham cohomology groups: 
\begin{equation*}
        H^{k}_{\text{dR}} (M, \mathbb{R} ) =  \text{ker } d_{k} /  \text{im } d_{k-1}    
\end{equation*}
The cup product in de Rham cohomology is exactly the exterior (or wedge) product on differential forms. 

De Rham's theorem states that the above groups are isomorphic to those derived via simplicial or singular cohomology \cite{BottTu}. 
And so we see that the topology of a manifold has a direct influence on the properties of differential forms that have it as their domain.  
For example, the Poincar\'e Lemma states that if $M$ is a contractible open subset of $\mathbb{R}^n$ then all smooth closed $k$-forms on $M$ are exact (the cohomology groups are trivial). 
In the language of multivariable calculus this becomes Helmholtz' theorem that a vector field, 
$\mathbf{V}$, with  $\text{curl} \mathbf{V} = 0$ in a simply connected open subset of $\mathbb{R}^3$ can be expressed as the gradient of a potential function:  $\mathbf{V} = \text{grad} f$ in the appropriate domain~\cite{NashSen}.  
These considerations play a key role in the study of electrodynamics via Maxwell's equations~\cite{GrossKotiuga}.

\section{Morse theory}
\label{MorseTheory}

We now turn to a primary topic in differential topology: to examine the relationship between the topology of a manifold $M$ and real-valued functions defined on $M$.   
The basic approach of Morse theory is to use the \textbf{level cuts} of a function $f: M \to \mathbb{R}$ and study how these subsets $M_a = f^{-1}(-\infty, a] $ change as $a$ is varied.  
For `nice' functions the level cuts change their topology in a well-defined way only at the critical points.  
This leads to a number of powerful theorems that relate the homology of a manifold to the critical points of a function defined on it.

\subsection{Basic results}

\begin{figure}
\centering
\includegraphics[width=0.4\textwidth]{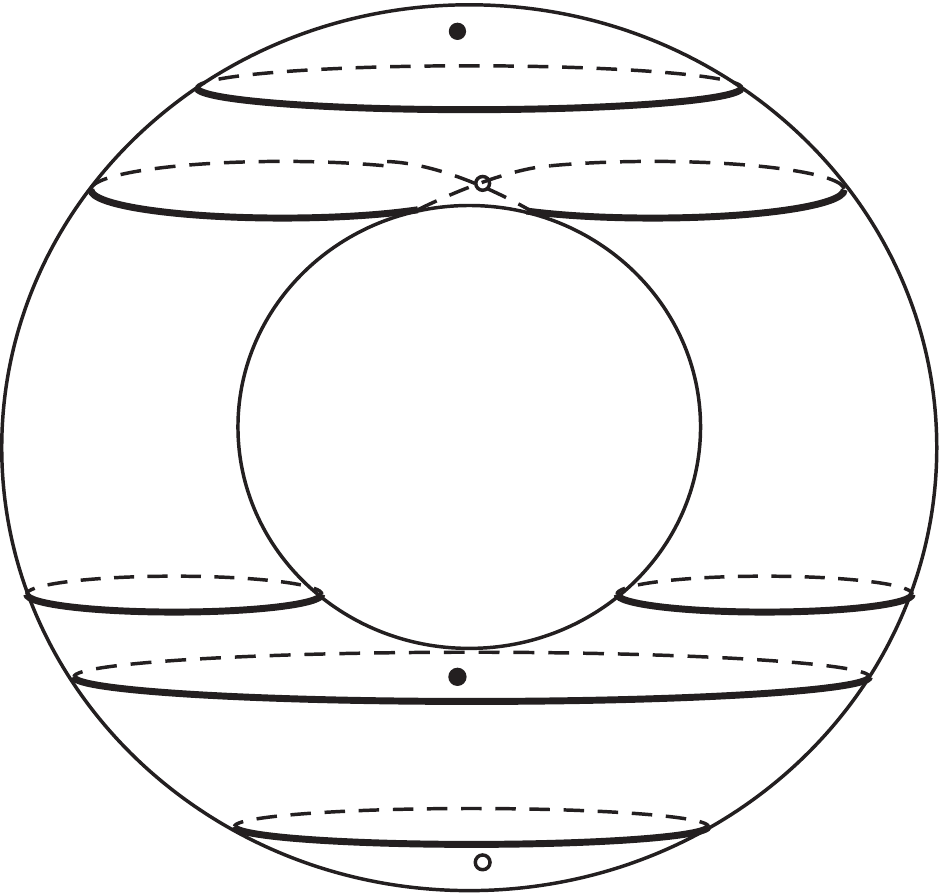}
\hspace{0.1\textwidth}
\includegraphics[width=0.4\textwidth]{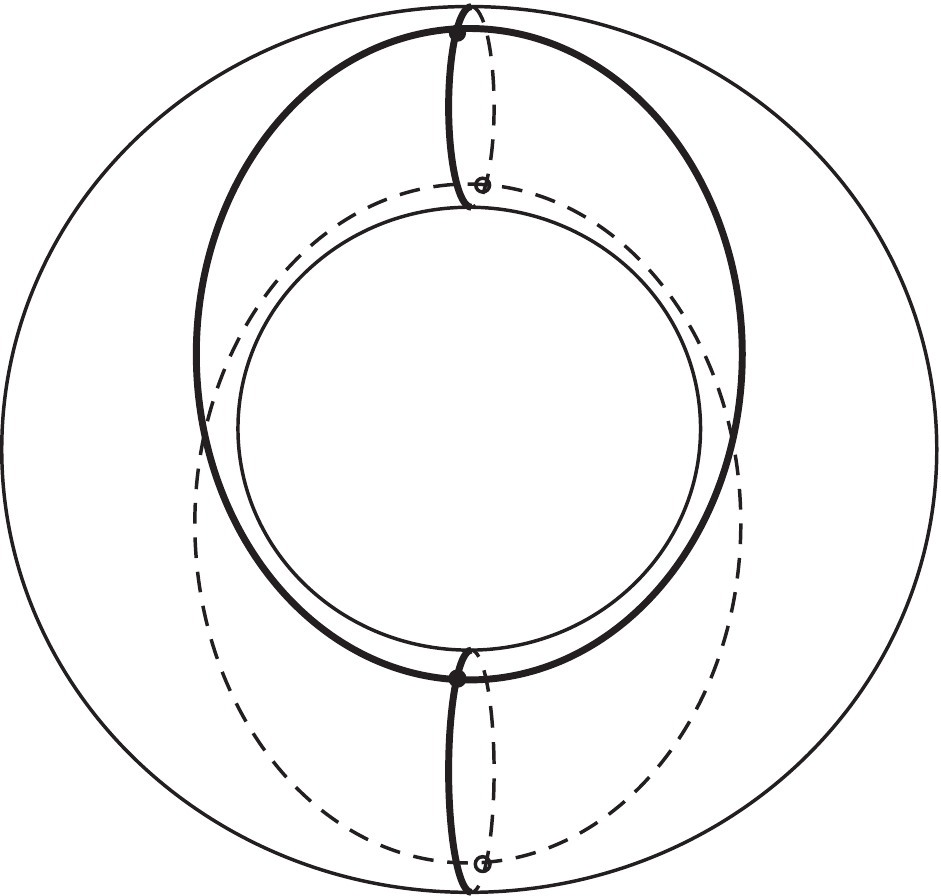}
\caption{Imagine a torus sitting with one point in contact with a plane and tilted slightly into the page as depicted. Define a Morse function by mapping each point on the torus to its height above the plane.  This function has four critical points: a minimum, two saddles and a maximum.  
\textbf{Left}: Five level cuts of the height function showing how the topology of a level cut changes when passing through a critical point.  
\textbf{Right}: Gradient flow lines between the maximum and the two saddle points, and from each saddle point to the minimum.  
}
\label{fig:morse_torus}
\end{figure}

A \textbf{Morse function} $f: M \to \mathbb{R}$ is a smooth real-valued function defined on a differentiable manifold $M$ such that each critical point of $f$ is isolated and the matrix of second derivatives (the \textbf{Hessian}) is non-degenerate at each critical point.  An example is illustrated in Fig.~\ref{fig:morse_torus}. 
The details on how to define these derivatives with respect to a coordinate chart on $M$ are given in Chapter~\ref{DifferentialGeometry}. 
This may seem like a restrictive class of functions but in fact Morse functions are dense in the space of all smooth functions, so any smooth function can be smoothly perturbed to obtain a Morse function \cite{Matsumoto:02}. 
Now suppose $x \in M$ is a critical point of $f$, i.e. $d f (x) = 0$.  
The \textbf{index} of this critical point is the number of negative eigenvalues of the Hessian matrix. 
Intuitively this is the number of directions in which $f$ is decreasing:  a minimum has index 0, and a maximum has index $m$ where $m$ is the dimension of the manifold $M$.  
Critical points of intermediate index are called \textbf{saddles} since they have some increasing and some decreasing directions.  

The two main results about level cuts $M_a$ of a Morse function $f$ are that: 
\begin{itemize}
\item When $[a, b]$ is an interval for which there are no critical values of $f$ 
(i.e.~there is no $x \in f^{-1}([a,b])$ for which $df(x) = 0$) 
then $M_a$ and $M_b$ are homotopy equivalent.  
\item Let $x$ be a non-degenerate critical point of $f$ with index $i$, let $f(x) = c$ and let $\epsilon>0$ be such that $f^{-1}[c-\epsilon, c+\epsilon]$ is compact and contains no other critical points.  Then $M_{c+\epsilon}$ is homotopy equivalent to $M_{c-\epsilon}$ with an $i$-cell attached.  
\end{itemize}
(Recall that an $i$-cell is an $i$-dimensional unit ball and the attaching map glues the whole boundary of the $i$-cell continuously into the prior space). 
The proofs of these theorems rely on homotopies defined via the \emph{negative gradient flow} 
of $f$~\cite{Matsumoto:02}.

Gradient flow lines are another key ingredient of Morse theory and allow us to define a chain complex and to compute the homology of $M$.  
Each point $x \in M$ has a unique \textbf{flow line} or \textbf{integral path} 
$\gamma_x : \mathbb{R} \to M$ such that 
\begin{equation*}
\gamma_x (0) = x  \text{ and }  \frac{\partial \gamma_x(t)}{\partial\, t} = - \nabla{f}(\gamma_x(t)).
\end{equation*}
Taking the limit as $t \to \pm \infty$, each flow line converges to a \textbf{destination} and an \textbf{origin} critical point.  
The \textbf{unstable manifold} of a critical point $p$ with index $i$ is the set of all $x \in M$ that have $p$ as their origin; this set is homeomorphic to an open ball of dimension $i$.  
Correspondingly, the \textbf{stable manifold} is the set of all $x$ that have $p$ as their destination. 
For suitably `nice' functions the collection of unstable manifolds form a cell complex for the manifold $M$ \cite{Banyaga:04}.  

We can also define a more abstract chain complex which is sometimes referred to as the \emph{Morse-Smale-Witten complex} to reflect the history of its development. 
Let $C_i$ be the chain group derived from formal sums of critical points of index $i$.  
A boundary operator $\partial: C_i \to C_{i-1}$ is then defined by mapping $p \in C_i$ to a sum of critical points $\sum \alpha_j q_j \in C_{i-1}$ for which there is a flow line with $p$ as its origin and $q$ as its destination. The coefficients $\alpha_j$ of the $q_j$ in this boundary chain are the number of geometrically distinct flow lines that join $p$ and $q_j$ (one can either count mod 2 or keep track of orientations in a suitable manner).  It requires some effort to show that $\partial \partial = 0$ in this setting; see \cite{Banyaga:04} for details.\footnote{%
In 2D, think of the flow lines that join a single maximum, minimum pair.  In general, such a region is bounded by flow lines from the maximum to two saddles and from these saddles to the minimum. The boundary of the maximum contains these two saddles and their boundaries contain the minimum in oppositely induced orientations.}
\textbf{Morse homology} is the homology computed via this chain complex. 

For \emph{finite-dimensional compact manifolds} Morse homology is isomorphic to singular homology, and we obtain the \textbf{Morse inequalities} relating numbers of critical points of $f : M \to \mathbb{R}$ to the Betti numbers of $M$: 
\begin{align*}
  c_0 &\geq \beta_0 \\
  c_1 - c_0 &\geq \beta_1 - \beta_0 \\
  c_2 - c_1 + c_0 &\geq \beta_2 - \beta_1 + \beta_0 \\
               &  \vdots   \\
 \sum_{0\leq i \leq m}  (-1)^{m-i} c_i   & =  \sum_{0\leq i \leq m} (-1)^{m-i} \beta_i  = \chi(M) . 
\end{align*} 
where $c_i$ is the number of critical points of $f$ of index $i$ and $\beta_i$ is the $i$-th Betti number of $M$. 
Notice that the final relationship is an \emph{equality}; the alternating sum of numbers of critical points is the same as the Euler characteristic $\chi(M)$.  
It also follows from the above that $c_i \geq \beta_i$ for each $i$.

\subsection{Extensions and applications}

Morse theory is primarily used as a powerful tool to prove results in other settings.  
For example,  Morse obtained his results in order to prove the existence of closed geodesics on a Riemannian manifold~\cite{Morse34};  most famously, Morse theory forms the foundation of a proof due to Smale of the higher-dimensional Poincar\'e conjecture~\cite{Smale61}.  
Morse theory has been extended in many ways that relax conditions on the manifold or the function being studied~\cite{Bott88}.  We mention a few of the main generalisations here. 

A \textbf{Morse-Bott function} is one for which the critical points may now not be isolated and instead form a critical set that is a closed submanifold.  At the very simplest level for example, this lets us study the height function of a torus sitting flat on a table since the circle of points touching the table is critical~\cite{Bott54}. 

The \textbf{Conley index} from dynamical systems is a generalization of Morse theory to flows in a more general class than those generated by the gradient of a Morse function. 
For general flows, invariant sets are no longer single fixed points but may be periodic cycles or even fractal ``strange attractors''.  
In the Morse setting, the index is simply the dimension of the unstable manifold of the fixed point, but for general flows a more subtle construction is required.  
Conley's insight was that an isolated invariant set can be characterized by the flow near the boundary of a neighborhood of the set. 
The Conley index is then (roughly speaking) the homotopy type of such a neighborhood relative to its boundary.  For details see \cite{ConleyEaston71,Conley78,Mischaikow02}. 

Building on Conley's work and the Morse complex of critical points and connecting orbits, Floer created an infinite-dimensional version of Morse homology now called \textbf{Floer homology}~\cite{Banyaga:04}.  
This has various formulations which have been used to study problems in symplectic geometry (the geometry of Hamiltonian dynamical systems) and also the topology of 3- and 4-dimensional manifolds \cite{McDuff05}. 

There are a number of approaches adapting Morse theory to a discrete setting, of increasing importance in geometric modelling, image and data analysis, and quantum field theory.  
The approach due to Forman is summarized  in the following section.

\subsection{Forman's discrete Morse theory}

Discrete Morse theory is a combinatorial analogue of Morse Theory for functions defined on cell complexes.
Discrete Morse functions are not intended to be approximations to smooth Morse functions, but the theory developed in \cite{forman98,forman02} keeps much of the style and flavor of the standard results from smooth Morse theory. 

In keeping with earlier parts of this chapter, we will give definitions for a simplicial complex 
$\mathcal{C}$, but the theory holds for general CW-complexes with little modification. 
First recall that a simplex $\alpha$ is a \textbf{face} of another simplex $\beta$ if $\alpha \subset \beta$, in which case $\beta$ is called a \textbf{coface} of $\alpha$.
A function $f: \mathcal{C} \to \mathbb{R}$ that assigns a real number to each simplex in $\mathcal{C}$ is a \textbf{discrete Morse function} if for every $\alpha^{(p)} \in \mathcal{C}$, $f$ takes a value less than or equal to $f(\alpha)$ on at most one coface of $\alpha$ and takes a value greater than or equal to $f(\alpha)$ on at most one face of $\alpha$.
In other words,
\begin{equation*} \label{DMF1}
	\# \{ \beta^{(p+1)} > \alpha \ | \ f(\beta) \leq f(\alpha) \} \leq 1, 
\end{equation*}
and
\begin{equation*} \label{DMF2} 
	\# \{ \gamma^{(p-1)} < \alpha \ | \ f(\gamma) \geq f(\alpha) \} \leq 1,
\end{equation*}
where $\#$ denotes the number of elements in the set.  
A simplex $\alpha^{(p)}$ is \emph{critical} if all cofaces take strictly greater values and all faces are strictly lower.

A cell $\alpha$ can fail to be critical in two possible ways.  There can exist $\gamma < \alpha$ such that $f(\gamma) \geq f(\alpha)$, or there can exist $\beta > \alpha$ such that $f(\beta) \leq f(\alpha)$.  
Lemma 2.5 of \cite{forman98} shows that these two possibilities are exclusive: they cannot be true simultaneously for a given cell $\alpha$.  Thus each non-critical cell $\alpha$ may be paired either with a non-critical cell that is a coface of $\alpha$, or with a non-critical cell that is a face of $\alpha$. 

\begin{figure}
\centering
\includegraphics[width=0.6\textwidth]{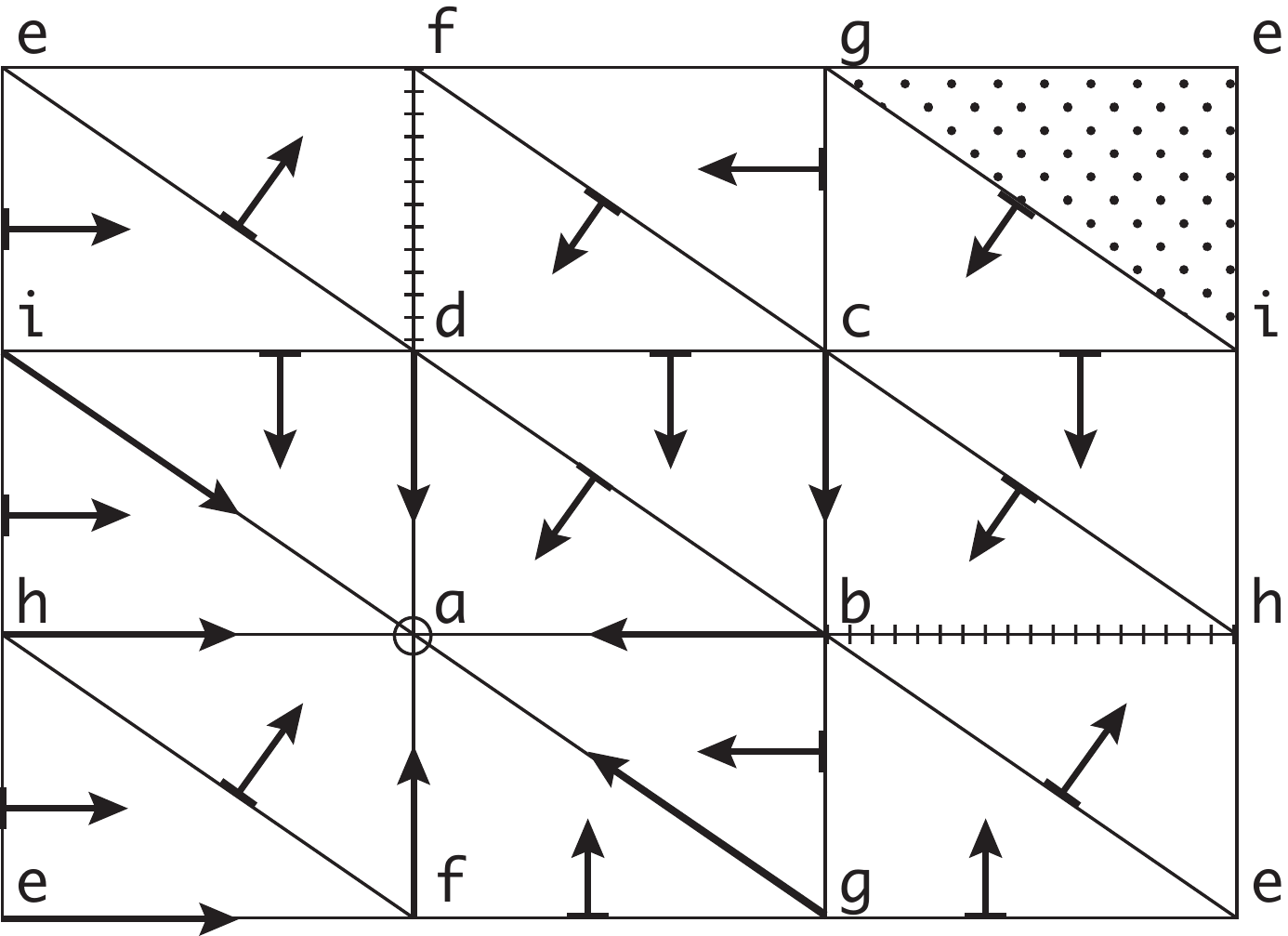}
\caption{ A simplicial complex with the topology of the torus (opposite edges of the rectangle are identified according to the vertex labels).  The arrows show how to pair simplices in a gradient vector field.  A compatible discrete Morse function has a critical 0-cell (a minimum) at $a$, two critical 1-cells (saddles) at edges $\langle b, h \rangle$ and $\langle d, f \rangle$ and a critical 2-cell (a maximum) at $\langle e, i, g \rangle$.  
}
\label{fig:discrete_morse}
\end{figure}

As noted by Forman (Section 3 of \cite{forman02}), it is usually simpler to work with pairings of cells with faces than to construct a discrete Morse function on a given complex. 
So we define a \textbf{discrete vector field} $V$ as a collection of pairs $( \alpha^{(p)}, \beta^{(p+1)})$ of cells $\alpha < \beta \in \mathcal{C}$ such that each cell of $\mathcal{C}$ is in at most one pair of $V$.  
A discrete Morse function defines a discrete vector field by pairing $\alpha^{(p)} < \beta^{(p+1)}$ whenever $f(\beta) \leq f(\alpha)$.
The critical cells are precisely those that do not appear in any pair. 
Discrete vector fields that arise from Morse functions are called \textbf{gradient} vector fields.  See Fig.~\ref{fig:discrete_morse} for an example. 

It is natural to consider the flow associated with a vector field and in the discrete setting the analogy of a flow-line is a $V$-path.
A \textbf{$V$-path} is a sequence of cells:
\begin{equation*}
    \alpha_0^{(p)}, \beta_0^{(p+1)},  \alpha_1^{(p)}, \beta_1^{(p+1)},  \alpha_2^{(p)}, \ldots, \beta_{r-1}^{(p+1)}, \alpha_r^{(p)}.    
\end{equation*}
where $(\alpha_i, \beta_i) \in V$, $\beta_{i}>\alpha_{i+1}$, and $\alpha_{i} \neq \alpha_{i+1}$ for all $i = 0,\ldots, r-1$.
A $V$-path is a \emph{non-trivial closed $V$-path} if $\alpha_r = \alpha_0$ for $r > 1$. Forman shows that a discrete vector field is the gradient vector field of a discrete Morse function if and only if there are no non-trivial closed $V$-paths (Theorem 9.3 of \cite{forman98}).

The four results about Morse functions that we gave earlier all carry over into the discrete setting:
the homotopy equivalence of level sets away from a critical point, adding a critical $i$-cell is homotopy equivalent to attaching an $i$-cell, the existence of and homology of the Morse chain complex, and the Morse inequalities.  
One of the notable differences between the discrete and continuous theories is that flow lines for a smooth Morse function on a manifold are uniquely determined at each point, whereas $V$-paths can merge and split.

\section{Computational topology}
\label{ComputationalTopology}

An algorithmic and combinatorial approach to topology has led to significant results in low-dimensional topology over the past twenty years.  
There are two main apsects to computational topology: first, research into methods for making topological concepts algorithmic, culminating for example, in the beginnings of an algorithmic classification of (Haken) 3-manifolds \cite{Matveev03} (a result analogous to the classification of closed compact 2-manifolds by Euler characteristic and orientability).  
And second, the challenge to find efficient and useful techniques for extracting topological invariants from data; see~\cite{edelsbook} for example.  
We start this section by describing simple algorithms that demonstrate the computability of the fundamental group and homology groups of a simplicial complex, and then survey some recent advances in building cell complexes and computing homology from data.   

\subsection{The fundamental group of a simplicial complex} 

In Section~\ref{Homotopy} we saw that the fundamental group of a topological space could be determined from unions and products of smaller spaces or by using a covering space.  When the space has a triangulation (i.e.~it is homeomorphic to a polyhedron) there is a more systematic and algorithmic approach to finding the fundamental group as the quotient of a free group by a set of relations that we summarize below.  See \cite{SpanierAT} for a complete treatment of this \textbf{edge-path group}.    

Let $\mathcal{C}$ be a connected finite simplicial complex.
Any path in $|\mathcal{C}|$ is homotopic to one that follows only edges in $\mathcal{C}$, and any homotopy between edge-paths can be restricted to the 2-simplices of $\mathcal{C}$.   
This means the fundamental group depends only on the topology of the 2-skeleton of $\mathcal{C}$.  
The algorithm for finding a presentation of $\pi_1(\mathcal{C})$ proceeds as follows. 

\begin{figure}
\centering
\includegraphics[width=0.6\textwidth]{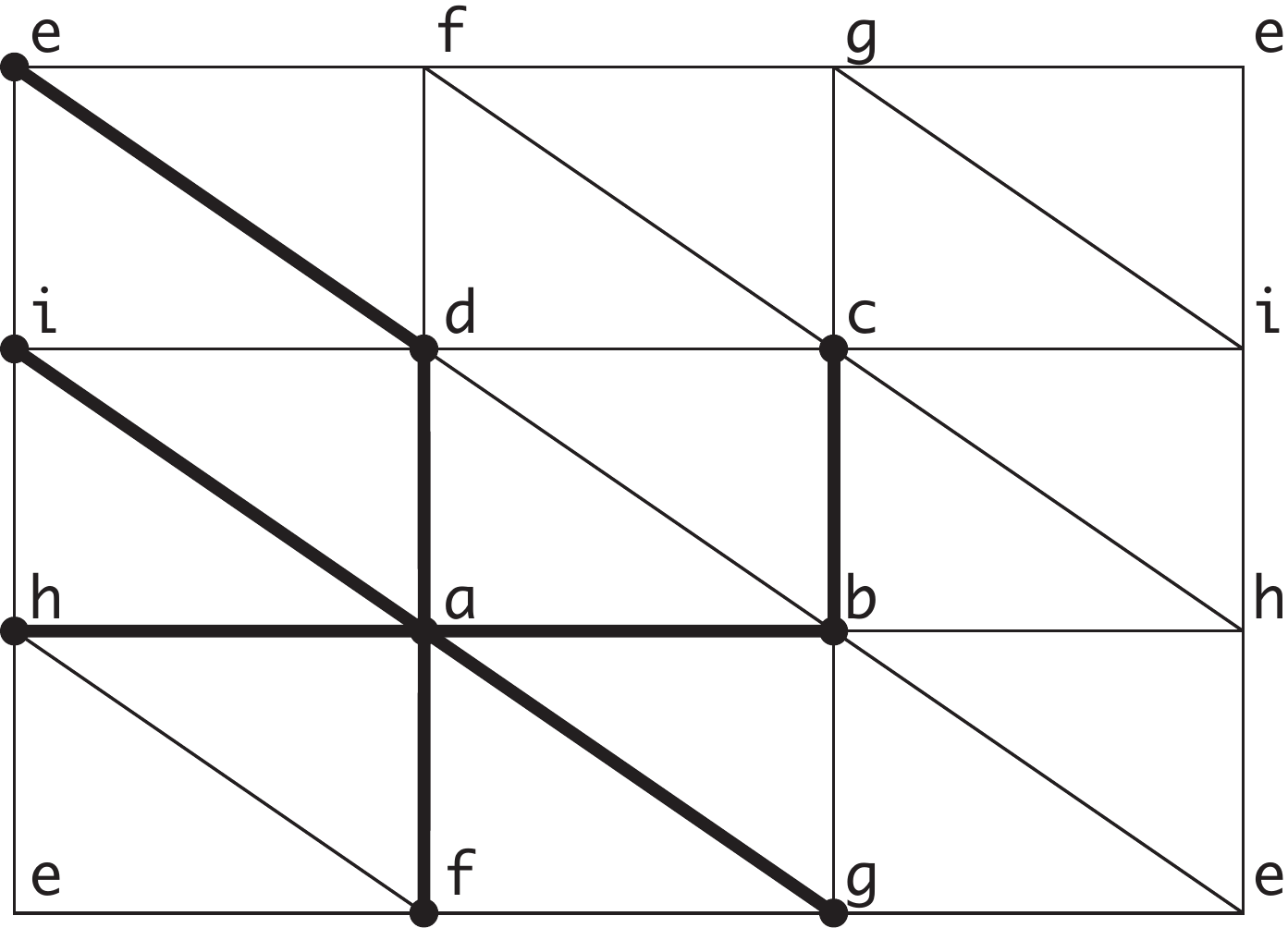}
\caption{A simplicial complex with the topology of a torus (opposite edges of the rectangle are identified according to the vertex labels).  A spanning tree $T$ with root vertex $a$ is shown in bold.  Any closed path that starts and ends at $a$ can be decomposed into a sum of loops that lie in $T$ except for a single edge.   
}
\label{fig:span_tree}
\end{figure}

First find a \textbf{spanning tree} $T \subset \mathcal{C}^{(1)}$ i.e.~a connected, contractible subgraph of the 1-skeleton that contains every vertex of $\mathcal{C}$; see Fig.~\ref{fig:span_tree} for an example.  
One algorithm for doing this simply grows from an initial vertex $v$ (\textbf{the root}) by adding adjacent (edge, vertex) pairs only if the other vertex is not already in $T$.  
A non-trivial closed edge-path in $\mathcal{C}$ (a loop) must include edges that are not in $T$  and in fact \emph{every} edge in $\mathcal{C} - T$ generates a \emph{distinct} closed path in 
$\mathcal{C}^{(1)}$.
Specifically, for each edge $\langle x_i,x_j \rangle \in \mathcal{C} - T$ there is a closed path starting and ending at the root $v$ and lying wholly in $T$ except for the generating edge; we label this closed path $g_{ij}$. 
Moreover, any closed path based at $v$ can be written as a concatenation of such generating paths where inverses are simply followed in the opposite direction: $g_{ji} =  g_{ij}^{-1}$.  
The $g_{ij}$ are therefore generators for a free group with coefficients in $\mathbb{Z}$.  
   
Next we use the 2-skeleton $\mathcal{C}^{(2)}$ to obtain the homotopy equivalences of closed edge-paths. Each triangle $\langle x_i, x_j, x_k \rangle \in \mathcal{C}$ defines a relation in the group via 
$g_{ij} g_{jk} g_{ki} = \text{id}$ (the identity) where we also set $g_{ij} = \text{id}$ 
if  $\langle x_i, x_j \rangle \in T$. 
Let $G(\mathcal{C}, T)$ be the finitely presented group defined by the above generators and relations.   
Then it is possible to show that we get isomorphic groups for different choices of $T$ and that $G(\mathcal{C}, T)$ is isomorphic to the fundamental group $\pi_1(|\mathcal{C}|)$~\cite{SpanierAT}.

If $\mathcal{C}$ has many vertices, then the presentation of its fundamental group as $G(\mathcal{C}, T)$ may not be a very efficient description.  
It is possible to reduce the number of generating edges and relations by using any connected and contractible subcomplex that contains all the vertices 
$\mathcal{C}^{(0)} \subset K \subset \mathcal{C}$.   
Generators for the edge-path group are labelled by edges in $\mathcal{C} - K$, and the homotopy relations are again defined by triangles in $\mathcal{C}^{(2)}$, but we can now ignore all triangles in $K$.  
For the example of the torus in Fig.~\ref{fig:span_tree} we could take $K$ to be the eight triangles in the $2 \times 2$ lower left corner of the rectangular grid.

\subsection{Smith normal form for homology} 

There is also a well defined algorithm for computing the homology groups from a simplicial complex $\mathcal{C}$.
This algorithm is based on finding the \emph{Smith normal form} (SNF) of a matrix representation of the boundary operator as outlined below.  

Recall that the oriented $k$-simplices form a basis for the $k$-th chain group, $C_{k}$.
This means it is possible to represent the boundary operator, $\partial_k: C_{k} \to C_{k-1}$, by a (non-square) matrix $A_k$ with entries in $\{-1,0,1\}$.  
The matrix $A_k$ has $m_k$ columns and $m_{k-1}$ rows where $m_{k}$ is the number of $k$-simplices in $\mathcal{C}$.  
The entry $a_{ij}$ is 1 if $\sigma_i \in C_{k-1}$ is a face of $\sigma_j \in C_k$ with consistent orientation, $-1$ if $\sigma_i$ appears in $\partial \sigma_j$ with opposite orientation and $0$ if $\sigma_i$ is not a face of $\sigma_j$.  
Thus each column of $A_k$ is a boundary chain  in $C_{k-1}$ with respect to a basis of simplices.  

The algorithm to reduce an integer matrix to SNF uses row and column operations as in standard Gaussian elimination, but at all stages the entries must remain integers. 
The row and column operations correspond to changing bases for $C_{k-1}$ and $C_{k}$ respectively and the resulting matrix has the form:
\begin{equation*}
\label{eq:SNF}
 D_k = \begin{bmatrix}  
       B_k & \mathbf{0} \\
       \mathbf{0} & \mathbf{0}
     \end{bmatrix} , 
     \quad \text{where} \quad
 B_k = \begin{bmatrix}  
       b_1 &       & 0 \\
           & \ddots &   \\
        0  &       & b_{l_k}
     \end{bmatrix} .
\end{equation*}
$B_k$ is a square matrix with $l_k$ non-zero diagonal entries that satisfy $b_i \geq 1$ and $b_1$ divides $b_2$, divides $b_3$, and so on. 
For a full description of the basic algorithm see Munkres \cite{MunkresAT}. 

The SNF matrices for $\partial_{k+1}$ and $\partial_{k}$ give a complete characterization of the $k$-th homology group $H_{k}$.
The rank of the boundary group $B_{k}$ ($\text{im }A_{k+1}$) is the number of non-zero rows of $D_{k+1}$, i.e., $l_{k+1}$. 
The rank of the cycle group $Z_{k}$ ($\text{ker }A_k$) is the number of zero columns of $D_{k}$, i.e.~$m_k - l_k$.
The torsion coefficients of $H_{k}$ are the diagonal entries $b_i$ of $D_{k+1}$ that are greater than one.
The $k$th Betti number is therefore 
\[ \beta_{k} = \text{rank}(Z_{k}) - \text{rank}(B_{k}) = m_k - l_k - l_{k+1}. \]  
Bases for $Z_{k}$ and $B_{k}$ (and hence $H_{k}$) are determined by the row and column operations used in the SNF reduction but the cycles found in this way typically have poor geometric properties.  

There are two practical problems with the algorithm for reducing a matrix to SNF as it is described in Munkres \cite{MunkresAT}.
First, the time-cost of the algorithm is of a high polynomial degree in the number of simplices; second, the entries of the intermediate matrices can become extremely large and create
numerical problems, even when the initial matrix and final normal form have small integer entries.  
When only the Betti numbers are required, it is possible to do better. 
In fact, if we construct the homology groups over the rationals, rather than the integers, then we need only apply Gaussian elimination to diagonalize the boundary operator matrices; doing this means we lose all information about the torsion however. 
Devising algorithms that overcome these problems and are fast enough to be effective on large complexes is an area of active research.

\subsection{Persistent homology}

The concept of persistent homology arose in the late 1990s from attempts to extract meaningful topological information from data \cite{Robins99,Frosini99,ELZ02}.
To give a finite set of points some interesting topological structure requires the introduction of a parameter to define which points are connected.    
The key lesson learnt from examining data was that rather than attempting to choose a single best parameter value, it is much more valuable to investigate a range of parameter values and describe how the topology changes with this parameter.  
So persistent homology tracks the topological properties of a sequence of nested spaces called a \textbf{filtration}    $  \cdots \subset \mathcal{C}_a \subset \mathcal{C}_b \subset \cdots  $
where $a < b \in \mathcal{I}$ is an index parameter. 
In a continuous setting, the nested spaces might be the level cuts of a Morse function on a manifold, so that $\mathcal{I}$ is a real interval. 
In a discrete setting this becomes a sequence of subcomplexes indexed by a finite set of integers. 
In either case as the filtration grows, topological features appear and may later disappear.  
The \textbf{persistent homology group}, $H_k(a,b)$ measures the topological features from 
$\mathcal{C}_a$ that are still present in $\mathcal{C}_b$. 
Formally, $H_k(a,b)$ is the image of the map induced on homology by the simple inclusion of $\mathcal{C}_a$ into $\mathcal{C}_b$.  
Algebraically, it is defined by considering cycles in $\mathcal{C}_a$ to be equivalent with respect to the boundaries in $\mathcal{C}_b$:    
\begin{equation*}
H_k(a, b) = Z_k(a) /  \left( B_k(b)  \cap Z_k(a) \right) . 
\end{equation*} 

Computationally, persistent homology tracks the birth and death of every equivalence class of cycle and provides a complete picture of the topological structure present at all stages of the filtration.  
The initial algorithm for doing this, due to Edelsbrunner, Letscher and Zomorodian \cite{ELZ02}, is surprisingly simple and rests on the observation that if we build a cell  complex by adding a single cell at each step, then (since all its faces must already be present) this cell either creates a new cycle and is flagged as positive,  or `fills in' a cycle that already existed and is labelled negative.  
If $\sigma$ is a negative $(k+1)$-cell, its boundary $\partial \sigma$ is a $k$-cycle and its cells are already flagged as either positive or negative.  
The new cell $\sigma$ is then paired with the most recently added (i.e.~youngest) unpaired positive cell in $\partial \sigma$.  If there are no unpaired positive cells available, we must grow $\partial \sigma$ to successively larger homologous cycles until an unpaired positive cell is found.  By doing this carefully we can guarantee that $\sigma$ is paired with the positive $k$-cell that created the homology class of $\partial \sigma$. 
Determining whether a cell is positive or negative \emph{a priori} is computationally non-trivial in general but there is a more recent version of the persistence pairing algorithm due to Zomorodian and Carlsson \cite{Zom05,Zom_review} that avoids doing this as a separate step, and also finds a representative $k$-cycle for each homology class.  

The result of computing persistent homology from a finite filtration is a list of pairs of simplices $(\sigma^{(k)}, \tau^{(k+1)})$ that represent the birth and death of each homology class in the filtration.  
The persistence interval for each homology class is then given by the indices at which the creator $\sigma$ and destroyer $\tau$ entered the filtration.  
Some non-trivial homology classes may be present at the final step of the filtration, these have an empty partner and are assigned `infinite' persistence.  
There are a number of ways to represent this persistence information graphically: the two most popular techniques are the \emph{barcode} \cite{Carlsson_barcode} and the \emph{persistence diagram} \cite{ELZ02}.  
The barcode has a horizontal axis representing the filtration index; for each homology class a solid line spanning the persistence interval is drawn in a stack above the axis. 
The persistence diagram plots the (birth, death) index pair for each cycle.  These points lie above the diagonal, and points close to the diagonal are homology classes that have low persistence. 
It is possible to show that persistence diagrams are stable with respect to small perturbations in the data.  Specifically, if the filtration is defined by the level cuts of a Morse function on a manifold, then a small perturbation to this function will produce a persistence diagram that is close to that of the original one \cite{CSEdelsHarer07}.

\subsection{Cell complexes from data} 

We now address how to build a cell complex and a filtration for use in persistent homology computations.    
Naturally, the techniques differ depending on the type of data being investigated; we discuss some  common scenarios below. 

The first construction is based on a general technique from topology called the \textbf{nerve of a cover}.  
Suppose we have a collection of `good' sets (the sets and their intersections should be contractible) 
 $\mathcal{U} = \{ U_1, \ldots, U_N \}$ whose union $\bigcup U_i$ is the space we are interested in. 
An abstract simplicial complex $\mathcal{N}(\mathcal{U})$ is defined by making each $U_i$ a vertex  and adding a $k$-simplex whenever the intersection $U_{i0} \cap \cdots \cap U_{ik} \neq \emptyset$. 
The nerve lemma states that $\mathcal{N}(\mathcal{U})$ has the same homotopy type as 
$\bigcup U_i$ \cite{HatcherAT}. 

\begin{figure}
\centering
\includegraphics[width=0.4\textwidth]{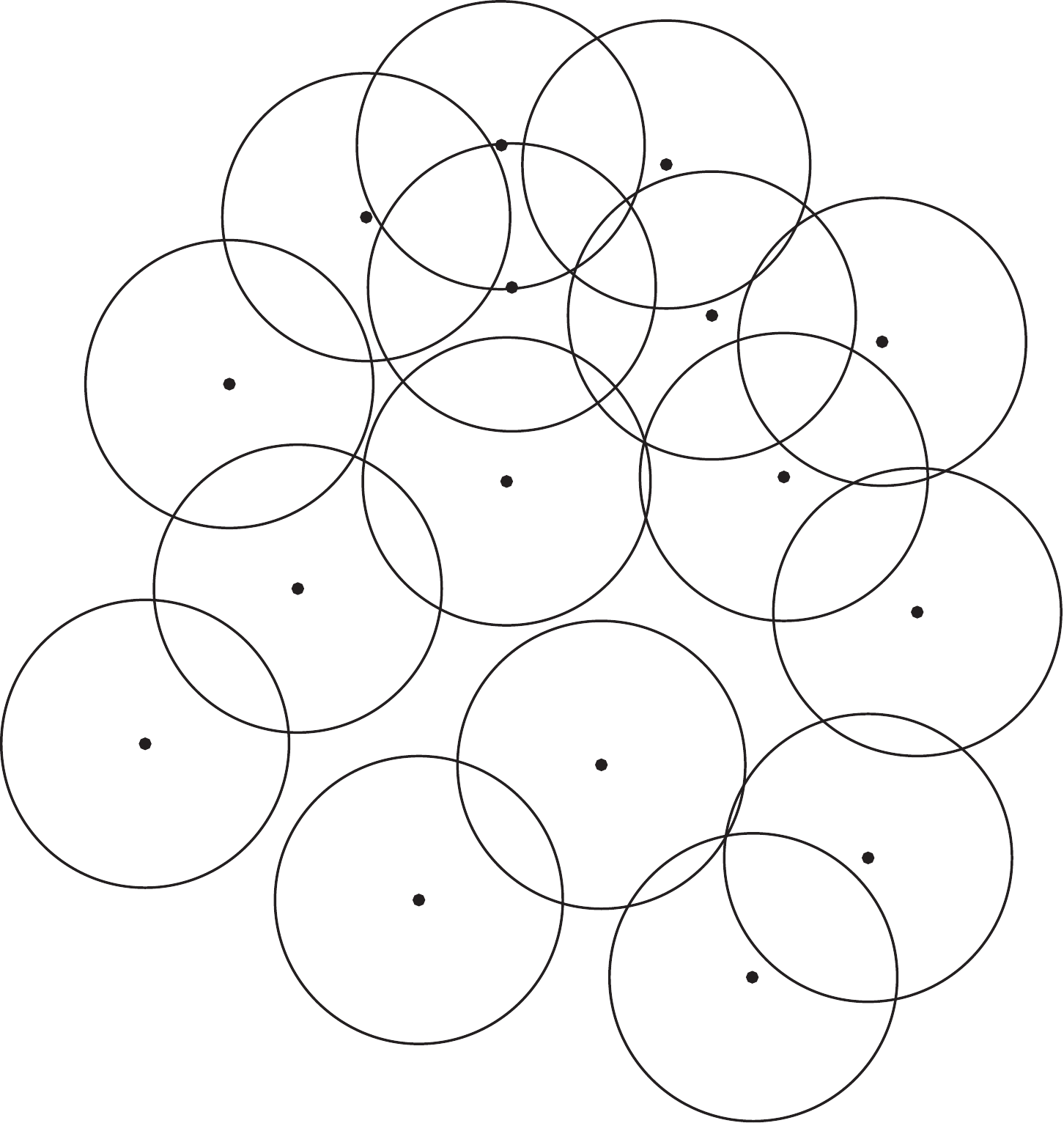}
\hspace{0.1\textwidth}
\includegraphics[width=0.4\textwidth]{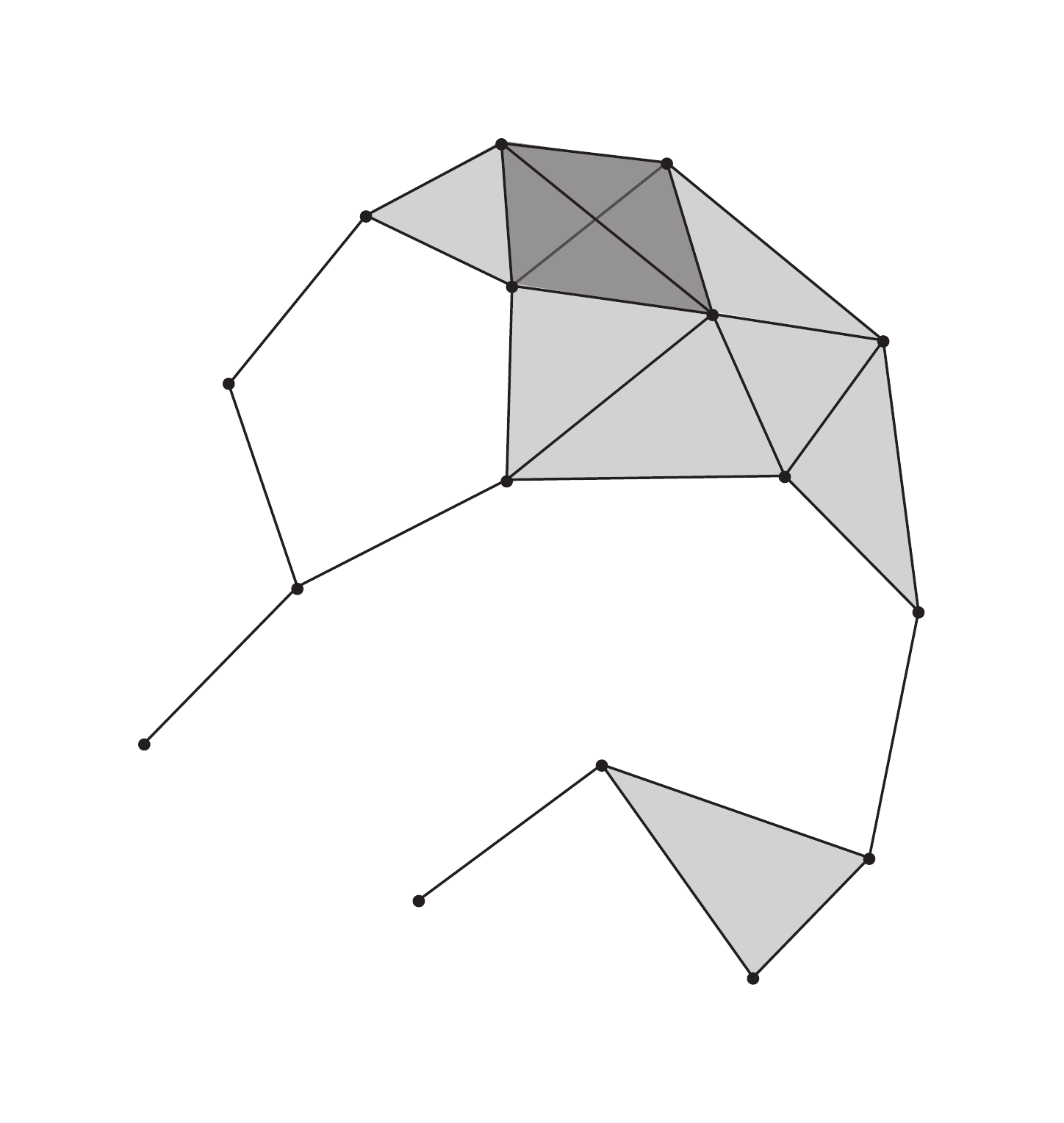}
\caption{\textbf{Left}: Balls of radius $a$ centered on 16 data points. \textbf{Right}: The nerve of the cover by balls of radius $a$ gives the \v{C}ech complex.  In this example the complex consists of points, edges, triangles, and a single tetrahedron (shaded dark gray).  As the radius of the balls increases there are more intersections between them and higher-dimensional simplices are created.  
}
\label{fig:cech}
\end{figure}

If the data set, $X$, is not too large, and the points are fairly evenly distributed over the object they  approximate, it makes sense to choose the $U_i$ to be balls of radius $a$ centered on each data point: $\mathcal{U}_{a} = \{ B(x_i, a), x_i \in X \}$.      
This is often called the \textbf{\v{C}ech complex}; see Fig.~\ref{fig:cech}. 
If $a < b$, we see that $\mathcal{N}(\mathcal{U}_{a}) \subset \mathcal{N}(\mathcal{U}_{b})$, and we have a filtration of simplicial complexes that captures the topology of the data as they are inflated from isolated points ($a = 0$) to filling all of space ($a \to \infty$).   

A similar construction to the \v{C}ech complex that is much simpler to compute is the \textbf{Vietoris-Rips} or \textbf{clique} complex.  Rather than checking for higher-order intersections of balls, we build a 1-skeleton from all pairwise intersections and then add a $k$-simplex when all its edges are present. 
This construction is not necessarily homotopy equivalent to the union of balls, but is useful when the data set comes from a high-dimensional space, perhaps with only an approximate metric.  

\begin{figure}
\centering
\includegraphics[width=0.32\textwidth]{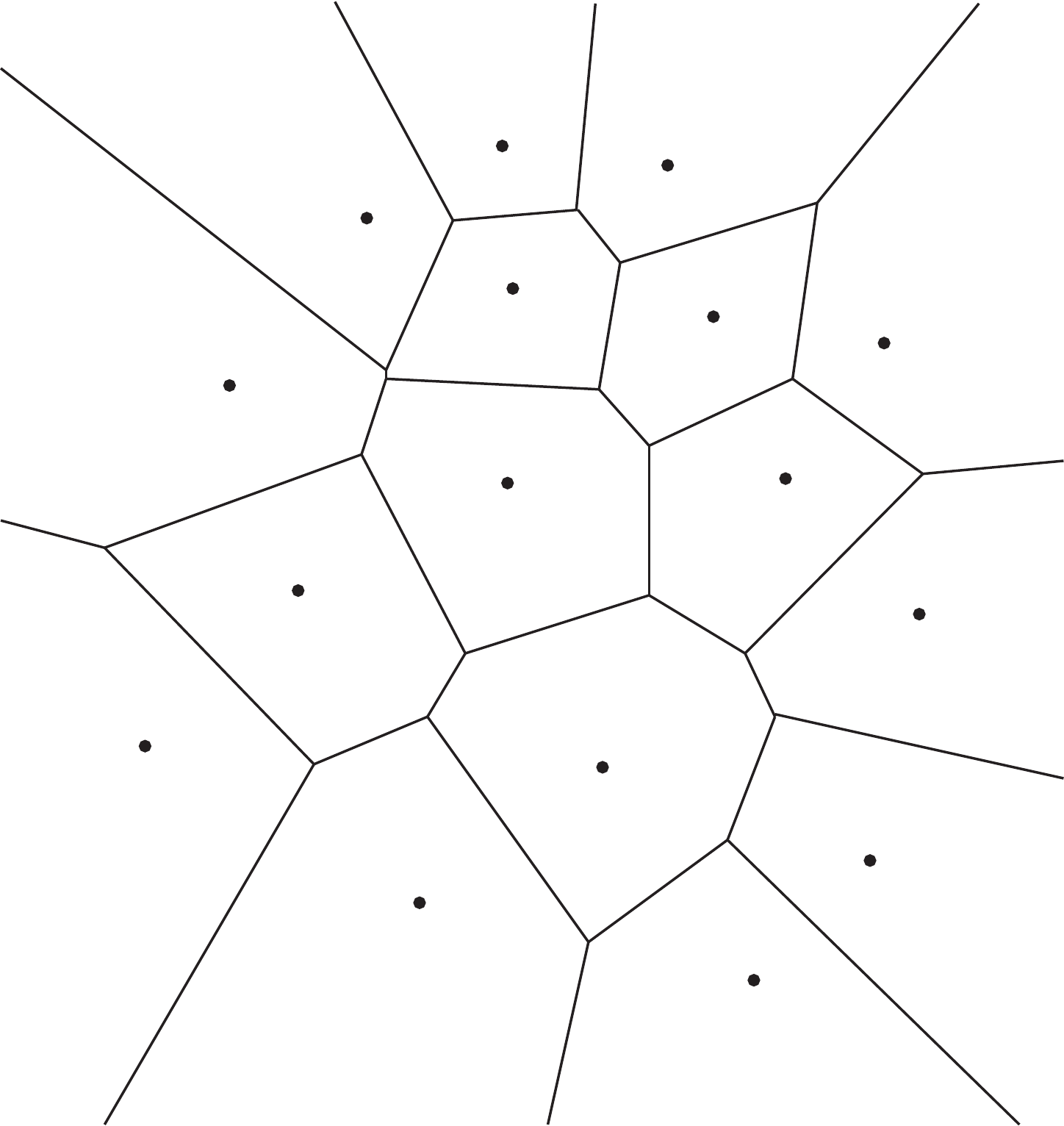}
\includegraphics[width=0.32\textwidth]{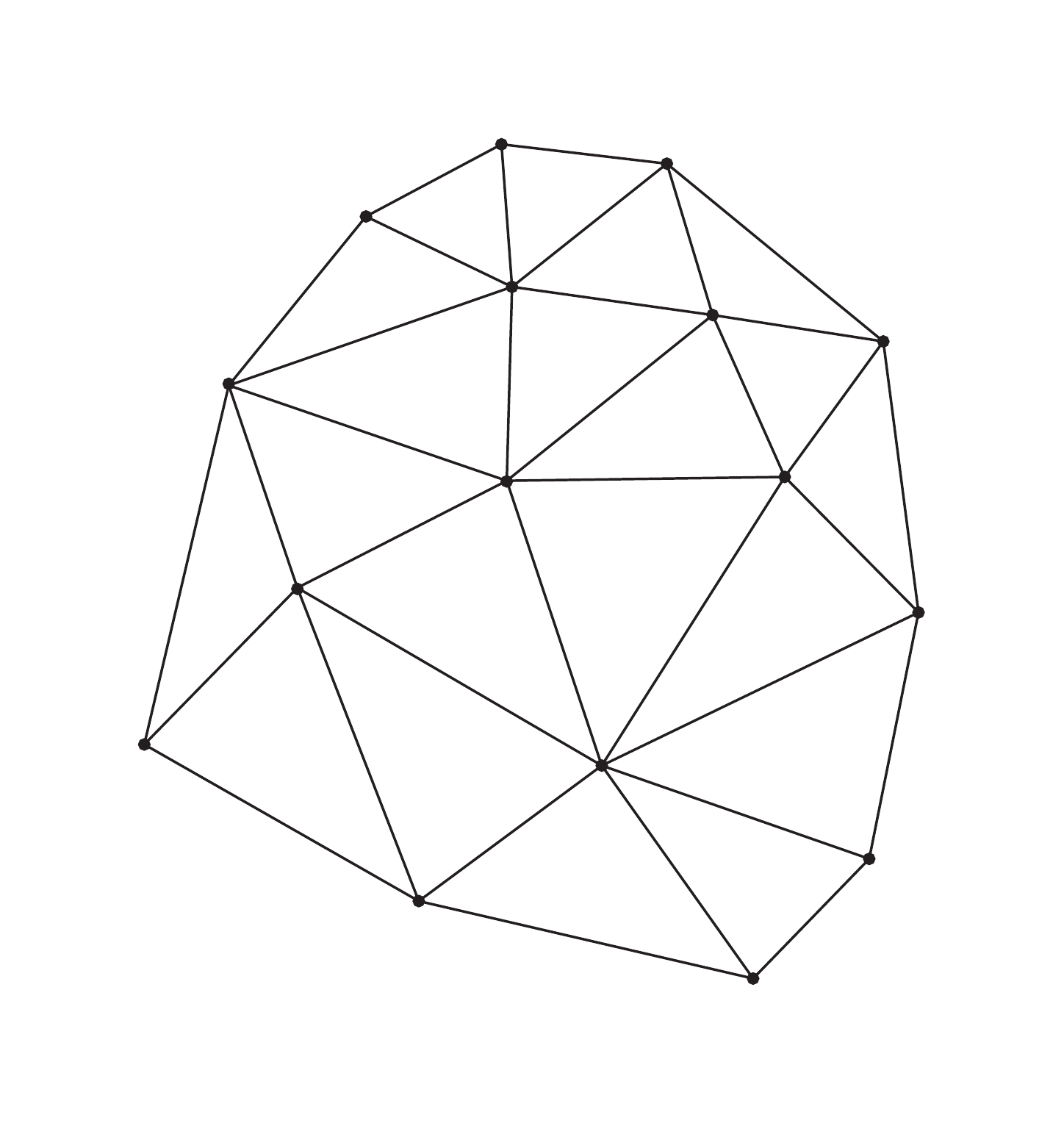}
\includegraphics[width=0.32\textwidth]{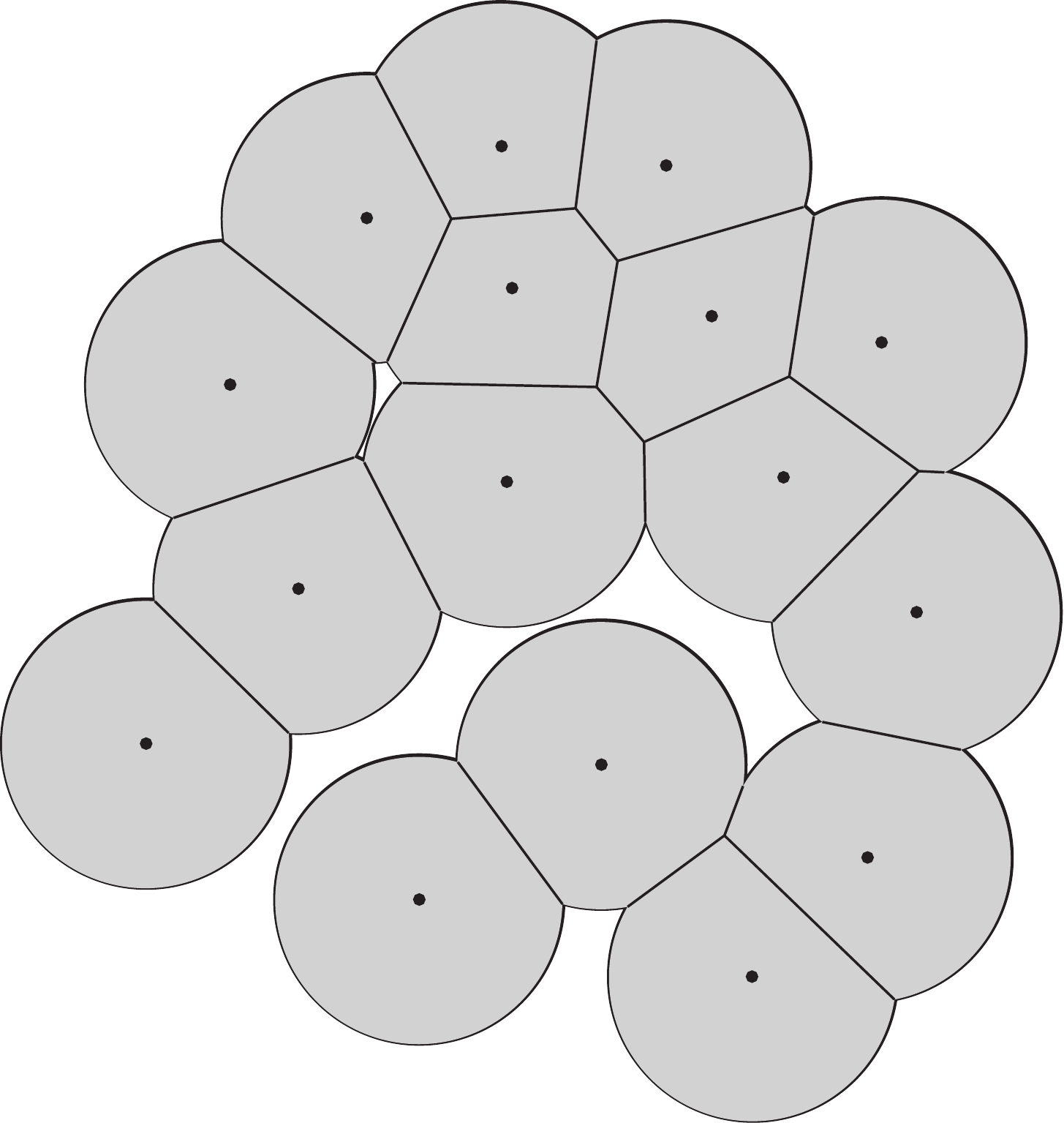}
\caption{\textbf{Left}: The Voronoi diagram of a data set with 16 points. \textbf{Centre}: The corresponding Delaunay triangulation.   \textbf{Right}: The union of balls of radius $a$ centred on the data points and partitioned by the Voronoi cells. The corresponding triangulation is almost the same as that shown in Fig.~\ref{fig:cech}: instead of the tetrahedron there are just two acute triangles. 
}
\label{fig:delaunay}
\end{figure}

A drawback of the \v{C}ech and Vietoris-Rips complexes is that many unnecessary high-dimensional simplices may be constructed.  
One way to avoid this is to build the Delaunay triangulation.  
There are many equivalent definitions of this widely-used geometric data structure \cite{SpatialTessellations}. 
We start by defining the \textbf{Voronoi partition} of space for a data set $\{ x_1, \ldots, x_N \} \subset \mathbb{R}^m$,  via the closed cells 
\begin{equation*}
 V( x_i) = \{ p \text{ such that } d(p, x_i) \leq d(p, x_j) \text{ for }  j \neq i \}. 
\end{equation*}
That is, the Voronoi cell of a data point is the region of space closer to it than to any other data point.  The boundary faces of Voronoi cells are pieces of the $(m-1)$-dimensional bisecting hyperplanes between pairs of data points.  
The \textbf{Delaunay complex} is the geometric dual to the Voronoi complex: when $k+1$ Voronoi cells share a $(m-k)$-dimensional face there is a $k$-simplex in the Delaunay complex that spans the corresponding $k+1$ data points. See Fig.~\ref{fig:delaunay} for an example in the plane ($m=2$).  
The geometry of the Voronoi partition guarantees that there are no simplices of dimension greater than $m$ in the Delaunay complex.  
\footnote{This is true for points in \emph{general position}.  Degenerate configurations of points occur, for example in the plane, when four Voronoi cells meet at a point.  In this case the Delaunay complex may be assigned either a 3-simplex, a quadrilateral cell, or one of two choices of triangle pairs.}

Now consider what happens when we take the intersection of each Voronoi cell with a ball centered on the data point, $B(x_i, a)$.  
The Voronoi cells partition the union of balls  $\bigcup B(x_i, a)$ and the geometric dual is a subset of the Delaunay complex that is commonly referred to as an \textbf{alpha complex} or alpha shape (where alpha refers to the radius of the ball \cite{Edels83,Edels94}). 
By increasing the ball radius from zero to some large enough value, we obtain a filtration of the Delaunay complex that starts with the finite set of data points and ends with the entire convex hull.  
The topology and geometry of alpha complexes has been used, for example, in characterizing the shape of and interactions between proteins \cite{Edels_bio_comp_top}.  
The Betti numbers of alpha shapes are also a useful tool for characterizing structural patterns of spatial data \cite{Robins06} such as the distribution of galaxies in the cosmic web \cite{vdW_etal_11}.

When the data set is very large, a dramatic reduction in the number of simplices used to build a complex is achieved by defining \emph{landmarks} and the \emph{witness complex}.  This construction generalises the Voronoi and Delaunay method, so that only a subset of data points (the landmarks) are used as vertices for the complex, whilst still maintaining topological accuracy.  A further advantage is that only the distances between data points are required to determine whether to include a simplex in the witness complex.  See \cite{carlsson04} for details, and \cite{carlsson09} for an extensive review of applications in data analysis.

Another important class of data is \textbf{digital images} which can be binary (voxels are black or white), greyscale (voxels take a range of discrete values), or coloured (voxels are assigned a multi-dimensional value).  
In this setting, the structures of interest arise from level cuts of functions defined on a regular grid.  
Morse theory is the natural tool to apply here, although in this application, the structures of interest are the level cuts of the function while the domain (a rectangular box) is simple. 
There are a number of different approaches to computing homology from such data and this is an area of active research.  
The works \cite{kaczynskibook,RWS11,Bendich10} present solutions motivated by applications in the physical sciences.

\section*{Guide to further reading}

We give a brief precis of a few standard texts on algebraic topology from mathematical and physical perspectives.  

Allen Hatcher's \textit{Algebraic Topology}~\cite{HatcherAT} is one of the most widely used texts in mathematics courses today and has a strong geometric emphasis.  
Munk\-res'~\cite{MunkresAT} is an older text that remains popular.  
Spanier~\cite{SpanierAT} is a dense mathematical reference and has one of the most complete treatments of the fundamentals of algebraic topology.  
A readable introduction to Morse theory is given by Matsumoto~\cite{Matsumoto:02} and Forman's review article~\cite{forman02} is an excellent introduction to his discrete Morse theory.  

Textbooks written for physicists that cover algebraic topology include Nakahara's comprehensive book \textit{Geometry, Topology and Physics}~\cite{Nakahara}, Schwarz \textit{Topology for Physicists}~\cite{Schwarz} and Naber \textit{Topology, Geometry and Gauge Theory}~\cite{Naber}.  
Each book goes well beyond algebraic topology to study its interactions with differential geometry and functional analysis.  A celebrated example of this is the Atiyah-Singer index theorem which relates the analytic index of an elliptic differential operator on a compact manifold to a topological index of that manifold, a result that has been useful in the theoretical physics of fundamental particles.



\bibliographystyle{plain}
\bibliography{arxiv_version}


\end{document}